\documentclass[11pt]{article}
\usepackage{amsmath, amssymb, graphicx, dsfont, xcolor, inputenc, natbib, enumerate, setspace}
\usepackage[margin=.75in]{geometry}
\DeclareMathOperator*{\argmax}{argmax}
\DeclareMathOperator*{\argmin}{argmin}

\newcommand{\Gn}{\mathbb{G}_n}
\newcommand{\Gzero}{\mathbb{G}_{n,0}}
\newcommand{\G}{\mathbb{G}}

\newtheorem{theorem}{Theorem}

\newtheorem{result}{Result}
\newtheorem{lemma}{Lemma}

\usepackage{soul}

\begin{document}

\title{Inference on function-valued parameters\\ using a restricted score test}
\author{Aaron Hudson, Marco Carone \& Ali Shojaie \\ University of Washington}
\date{}
\setlength{\abovedisplayskip}{0pt}
\setlength{\belowdisplayskip}{0pt}

\maketitle

\begin{abstract}
It is often of interest to make inference on an unknown function that is a local parameter of the data-generating mechanism, such as a density or regression function. Such estimands can typically only be estimated at a slower-than-parametric rate in nonparametric and semiparametric models, and performing calibrated inference can be challenging. In many cases, these estimands can be expressed as the minimizer of a population risk functional. Here, we propose a general framework that leverages such representation and provides a nonparametric extension of the score test for inference on an infinite-dimensional risk minimizer. 
We demonstrate that our framework is applicable in a wide variety of  problems. 
As both analytic and computational examples, we describe how to use our general approach for inference on a mean regression function under (i) nonparametric and (ii) partially additive models, and evaluate the operating characteristics of the resulting procedures via simulations.
\end{abstract}

\section{Introduction}

It is often the case that the estimand of scientific interest in a given application is an unknown function, either in its entirety or through its evaluation at one or several points in its domain. As a parameter of the underlying data-generating mechanism, such function is typically either global --- such as a distribution or quantile function --- or local --- such as a density, conditional mean or hazard function. When sufficiently strong parametric assumptions are made, inference on such an estimand, be it local or global, is usually straightforward, and parametric rates of estimation are achievable. For instance,  it is common to assume that a regression function is linear or  polynomial of a fixed degree, and inference then only involves a finite collection of unknown regression coefficients, which can be readily estimated at the parametric rate under weak conditions. However, such restrictive modeling assumptions bear the risk of invalid inference due to model misspecification. This fact has motivated investigators to instead rely on nonparametric or semiparametric models, for which the risk of model misspecification is reduced.

In nonparametric and semiparametric models, whether the parameter is local or global determines whether regular parametric-rate estimators exist for the unknown function, and thereby, how challenging calibrated inference is to achieve. When the unknown function is a global parameter of the data-generating mechanism,  parametric-rate inference is possible, and in fact, there is a well-established efficiency theory that characterizes the large-sample behavior of optimal estimators \citep{bickel1998efficient}. There also exist several constructive approaches for obtaining such estimators, including one-step debiasing procedures \citep{pfanzagl1982contributions}, estimating equations \citep{van2003unified, chernozhukov2018double}, and targeted minimum loss-based estimation \citep{van2011targeted}. This contrasts sharply with situations in which the unknown function is a local parameter of the data-generating mechanism.

When the unknown function is a local parameter of the data-generating mechanism, there usually does not exist any regular parametric-rate estimator. While estimation strategies abound for this setting, there are relatively fewer formal approaches for inference (e.g., construction of confidence sets and hypothesis tests) based on these strategies since studying the limiting distribution of such estimators is usually challenging.  For example, it is often the case that the bias of an estimator obtained by minimizing an empirical risk criterion tends to zero at the same rate as its standard error. To address this bias and facilitate inference, several approaches have been proposed, most commonly in the context of kernel smoothing.
One common approach consists of constructing a data-driven bias correction \citep[see, e.g.,][]{hardle1991bootstrap, eubank1993confidence, sun1994simultaneous, calonico2018effect, lu2020kernel}. Another approach consists of undersmoothing, that is, selecting tuning parameter values that deliberately inflate the variance in order to deflate the bias of the estimator, even though such choice results in a suboptimal risk value \citep[see, e.g.,][]{hall1991edgeworth, hall1992effect, neumann1995automatic}.
Both approaches typically require a characterization of the bias of the considered estimator, which often has a complex form and is difficult to estimate. They can also be sensitive to tuning parameter selection and difficult to implement in practice. As an alternative, \cite{hall2013simple} suggests a  bootstrap-based algorithm for pointwise inference, whereas \cite{van2018cv} proposes to use targeted minimum loss-based estimation on a  sequence of decreasingly-regularized modifications of the original estimand. However, neither approach appears to be directly applicable when uniform coverage or simultaneous testing is of interest.

In many settings, the estimand of interest can be represented as the minimizer of a population risk functional. Here, we leverage this common representation to develop a novel general framework for inference for function-valued parameters in nonparametric and semiparametric models. We propose a test based on assessing the feasibility of the null parameter value by evaluating the extent to which the derivative of the risk functional at this value appears to equal zero. Our test is an infinite-dimensional extension of the classical score test of \cite{rao1948large}, and in fact, reduces to this test when the model considered is finite-dimensional. By inverting the proposed test, we obtain uniform confidence bands for the unknown function of interest or its evaluation on a set. In contrast to existing approaches applicable in infinite-dimensional models, our proposal does not require estimation of the unknown function itself, thereby circumventing the difficulties usually caused by the bias of existing estimators. The framework we propose is quite general, and applies equally to classical parameters, such as density and conditional mean functions, as to more complicated parameters, which we discuss later in the paper, so long as the parameter of interest is a population risk minimizer. Our proposal is also flexible, requiring few assumptions about the data-generating mechanism.


The rest of the paper is organized as follows. In Section~\ref{sec:overview}, we present a high-level sketch of our proposed framework for inference. In Section~\ref{sec:derest}, we discuss estimation of risk functional derivatives, which play a critical role in our proposed approach. We present the theoretical results supporting our approach in Section~\ref{sec:properties}, and discuss practical considerations arising in its implementation in Section~\ref{sec:implementation}. 
We evaluate the operating characteristics of the proposed method in Section~\ref{sec:sims}, and apply it to data from the 1987 National Medical Expenditures Survey in Section~\ref{sec:realdata}. We provide concluding remarks in Section~\ref{sec:disc}.

\section{Overview of the proposed framework}\label{sec:overview}

\subsection{Preliminaries}

We begin by introducing some definitions and the notation used throughout.
Let $Z_1,Z_2,\ldots,Z_n$ represent independent random vectors drawn from a distribution $P_0$ known only to reside in a potentially rich statistical model $\mathcal{M}$, and denote by $\mathcal{Z}$ the sample space corresponding to $P_0$.
Suppose that $\Theta$ is a given function class and $P\mapsto\theta_P\in\Theta$ is a function-valued parameter mapping defined over $\mathcal{M}$. We are interested in making inference on $\theta_0:=\theta_{P_0}$. Suppose that for each $P\in\mathcal{M}$ there exists a $P$-risk functional $R_{P}:\Theta\rightarrow\mathbb{R}$ such that $\theta_P =\argmin_{\theta \in \Theta}R_{P}(\theta)$. In particular, defining the shorthand notation $R_0:=R_{P_0}$, this allows us to write the representation $\theta_0=\argmin_{\theta\in\Theta}R_0(\theta)$. In many cases, the risk functional has the simpler form \begin{align}\label{expectation}
R_P(\theta)=E_P[\ell_{P}(\theta,Z)]
\end{align} for some loss function $\ell_{P}:\Theta\times\mathcal{Z}\rightarrow\mathbb{R}$ indexed by $P$, though this is never required below.

\subsection{Working examples}

Minimizers of risk functionals arise naturally in many problems. Before describing our proposed approach, we describe the two working examples we will refer to throughout the manuscript. Both examples pertain to conditional mean functions. Let $Z := (X,W, Y)$, where $Y\in\mathbb{R}$ is the response variable and $(X,W)\in\mathbb{R}\times \mathbb{R}^d$ represents a covariate vector.
In the first example,
we will consider inference on the conditional mean function $\theta_0:x\mapsto E_0\left(Y\,|\,X=x\right)$ under a nonparametric model, where here and below $E_0$ denotes expectation under $P_0$.
Because we can express $\theta_0$ as the minimizer of the least-squares risk, that is,
\begin{align}
\theta_0 =\argmin_{\theta\in\Theta}E_0\,[ \{Y - \theta(X)\}^2]
\label{RegressionRM}
\end{align}with $\Theta$ taken to be the space $L_2(P_0)$ of $P_0$-square-integrable real-valued functions defined on the support of $X$, this corresponds to using the $P$-risk functional $R_{P}(\theta) := E_P\,[\{Y - \theta(X)\}^2]$. We readily see that $P\mapsto R_P(\theta)$ is a linear functional.  In fact, this risk functional has the form \eqref{expectation} with $\ell_P:(\theta,z)\mapsto \{y-\theta(x)\}^2$, where $\ell_P$ does not depend on $P$ at all.

In the second example, 
we will consider inference on $\theta_0$ under a partially additive mean model that enforces the structure 
$E_0\left(Y\,|\,X=x,W=w\right)=\theta_0(x)+f_0(w)$ for unknown functions $\theta_0$ and $f_0$ with condition $E_0\left[\theta_0(X)\right]=0$ imposed for identifiability.  The partially linear model is a special case of the partially additive model under which $\theta_0$ must also be linear \citep{robinson1988root}.
The parameter value $\theta_0$ facilitates a quantification of the association between an exposure $X$ and outcome $Y$ after adjustment for a vector $W$ of potential confounders: specifically, $\theta_0(x_1) - \theta_0(x_0)$ represents the difference in mean outcome between two subpopulations of individuals with exposure levels $x_1$ and $x_0$ but same level of confounding factors.
Similarly as before, $\theta_0$ can be expressed as a minimizer of a population least-squares risk, that is,
\begin{align}
\theta_0 = \argmin_{\theta\in\Theta}E_0\,[\{ Y - f_0(W) - \theta(X) \}^2]
\label{PartAddModRM}
\end{align}with $\Theta$ now taken to be the subset $L_2^0(P_0)$ of elements of $L_2(P_0)$ with $P_0$-mean zero. This corresponds to using the $P$-risk functional $R_P(\theta):=E_P\,[\{Y-f_P(W)-\theta(X)\}^2]$, where the nuisance function $f_P$ is such that $E_P\left(Y\,|\,X=x,W=w\right)=\theta_P(x)+f_P(w)$ for each $x$ and $w$. In this case, the risk functional also has the form \eqref{expectation} with $\ell_P:(\theta,z)\mapsto \{y-f_P(w)-\theta(x)\}^2$, a loss function that depends on $P$ through $f_P$. As such, in this case, $\theta\mapsto R_P(\theta)$ is not a linear functional.

\subsection{General inferential strategy}

Our objective is to conduct formal inference for an arbitrary population risk minimizer as defined above.
We begin by constructing a level $\alpha \in (0,1)$ test of the hypothesis
\begin{align*}
H_0: \theta_0 = \theta_*\ ,
\end{align*}
where $\theta_*\in\Theta$ is a pre-specified null parameter value, against the complement hypothesis $H_1:\theta_0\neq \theta_*$.
Then, by inverting this test, we derive a confidence region $\mathcal{C}_n$ for $\theta_0$, that is, we obtain a random set $\mathcal{C}_n=\mathcal{C}_n(Z_1,Z_2,\ldots,Z_n)\subset \Theta$ that contains the population risk minimizer with probability at least $1-\alpha$ as sample size $n$ tends to infinity:
\begin{align*}
\liminf_{n \to \infty} P_0\,(\theta_0 \in \mathcal{C}_n\,) \geq 1-\alpha\ .
\end{align*}

Our proposal is closely related to the classical (parametric) score test of \cite{rao1948large}, which we briefly review.
Suppose that $\Theta$ is the collection of linear functions $\left\{\theta_\beta:z\mapsto z^\top\beta: \beta \in \mathbb{R}^p\right\}$, a set indexed by the vector $\beta$ of finite dimension $p$.
In the finite-dimensional setting, for any null value $\beta_*\in\mathbb{R}^p$ of the index parameter, the classical score test assesses whether the null function  $\theta_{\beta_*}$  is a population risk minimizer by determining whether there is empirical evidence to suggest that the derivative of the risk function $\beta\mapsto R_{0}(\theta_\beta)$ evaluated at $\beta_*$ is zero. If the derivative is nonzero, $\theta_{\beta_*}$ cannot be a population risk minimizer.
This approach is based on studying local perturbations of $R_{0}$  in a neighborhood of the null value $\theta_{\beta_*}$ along a finite-dimensional collection of directions. As such, it does not rely upon estimation of the population risk minimizer $\theta_0$.
The score test is therefore appealing when it is difficult to construct an estimator of $\theta_0$ with a tractable limiting distribution.

Our proposal generalizes the classical score test to the infinite-dimensional setting.
For this generalization, we will require a proper notion of differentiability of $R_0$ over the infinite-dimensional space $\Theta$. For simplicity, suppose that $\Theta$ is a convex space. We will say that the population risk functional $R_0:\Theta\rightarrow\mathbb{R}$ is  G\^ateaux differentiable at $\theta=\theta_*$ provided that, for each direction $h\in\mathcal{H}(\theta_*):=\{\theta-\theta_*:\theta\in\Theta\}$, the G\^{a}teaux derivative
\begin{align*}
\dot{R}_{0,\theta_*}(h) := \lim_{c \to 0}\frac{R_{0}(\theta_* + ch) - R_{0}(\theta_*)}{c} = \left.\frac{d}{dc}R_{0}(\theta_* + ch)\right|_{c=0}
\end{align*} exists and is finite, and furthermore, the functional $\dot{R}_{0,\theta_*}:\mathcal{H}(\theta_*)\rightarrow\mathbb{R}$ is linear. The G\^ateaux derivative describes the rate at which the risk functional $R_{0}$ changes in value when making an infinitesimal shift away from $\theta_*$ in the direction $h$.
The key observation we use is that, since $\theta_0$ is an optimizer of $R_0$, the G\^ateaux derivative $\dot{R}_{0,\theta_0}(h)$ of $R_0$ at $\theta_0$ must be zero in any direction $h$, that is, $\dot{R}_{0,\theta_0}(h)=0$ for each $h\in\mathcal{H}(\theta_0)$. Thus, under $H_0: \theta_0 = \theta_*$, it must also be that $\dot{R}_{0,\theta_*}(h) = 0$ for each $h\in\mathcal{H}(\theta_*)$. Conversely, if $H_1$ is instead true, there must exist some function $h_*\in\mathcal{H}(\theta_*)$ such that $\dot{R}_{0,\theta_*}(h_*)\neq 0$.
To test $H_0$ against $H_1$, we assess the existence of such a direction $h_*$.

Formally, our objective can be restated as determining whether the G\^ateaux derivative of $R_0$ at $\theta_*$ in the steepest direction is zero, that is, we note that the null hypothesis $H_0$ can be reframed as
\begin{align*}
H_0:\sup_{h \in \mathcal{H}(\theta_*)} |\dot{R}_{0,\theta_*}(h)| = 0\ .
\end{align*}
If the function class $\Theta$ --- and therefore $\mathcal{H}(\theta_*)$ as well --- is rich, it may not be feasible to determine if $\dot{R}_{0,\theta_*}(h)\neq 0$ for any direction $h\in\mathcal{H}(\theta_*)$.
We may instead consider a subclass $\mathcal{H}\subseteq \mathcal{H}(\theta_*)$ of directions to investigate, and then assess whether there is empirical evidence to reject the restricted null hypothesis
\begin{align*}
H_{0,r}:\sup_{h \in \mathcal{H}} |\dot{R}_{0,\theta_*} (h)| = 0\ .
\end{align*}
Of course, if $H_{0,r}$ is not true, then neither is $H_0$, but the converse statement does not hold. Thus, a calibrated test of $H_{0,r}$ against its complement will generally constitute a conservative test of $H_0$ against its complement. We refer to our method as the \textit{restricted score test} because we aim assess $H_{0,r}$ against its complement $H_{1,r}$, that is, to determine if the  G\^ateaux derivative of greatest magnitude over the restricted space $\mathcal{H}$ is zero or not.

Our approach to assessing $H_{0,r}$ consists of measuring the aggregate `size' of  the collection of G\^ateaux derivatives evaluated at $\theta_*$ in each direction $h \in \mathcal{H}$.
To do so, we may select any norm $\Omega$ defined on the vector space $\ell^\infty(\mathcal{H})$ of bounded real-valued functionals on $\mathcal{H}$, and then use $\Omega(\dot{R}_{0,\theta_*})$ as a measure of departure from $H_{0,r}$.
Such an approach is valid because if $H_{0,r}$ holds, $\Omega(\dot{R}_{0,\theta_*})=0$.
We later show that, under appropriate conditions on $\mathcal{H}$, we can construct an estimator $\dot{R}_{n,\theta_*}$ of $\dot{R}_{0,\theta_*}$ such that, as a random element in $\ell^\infty(\mathcal{H})$,  the normalized process $
 \{ n^{1/2}[\dot{R}_{n,\theta_*}(h) - \dot{R}_{0,\theta_*}(h)]: h \in \mathcal{H} \}$ converges weakly to a tight mean-zero Gaussian process $\mathbb{G}:=\{\mathbb{G}(h):h\in\mathcal{H}\}$  relative to the supremum norm. Because $\dot{R}_{0,\theta_*}$ is the zero element whenever $\theta_0=\theta_*$, we may then test $H_{0,r}$ using the test statistic $\Omega(n^{1/2}\dot{R}_{n,\theta_*})$, which has a tractable limiting distribution that can be approximated using resampling techniques, as we demonstrate later.

We propose to construct confidence regions for $\theta_0$ by inverting our restricted score test.
Let $\Psi := \left\{\psi_u: u \in \mathcal{U}\right\}$ denote a collection of real-valued functionals on $\Theta$ indexed by some set $\mathcal{U}$.
We define $\mathcal{C}_{n}:=\{\theta\in\Theta:\text{we fail to reject $\theta_0=\theta$ against $\theta_0\neq \theta$ based on $Z_1,Z_2,\ldots,Z_n$}\}$ as the set of functions in $\Theta$ compatible with the available data and set  $\Psi_n(u):=\{\psi_u(\theta):\theta\in\mathcal{C}_{n}\}$.
Later, we show that a simultaneous confidence region for any smooth functional $\psi_u(\theta_0)$ is given by 
\begin{align*}
\left(\inf\Psi_n(u),\ \sup\Psi_n(u)\right)\ .
\end{align*} 
In particular, we can set  $\mathcal{U}$ to be some subset $\mathcal{O}$ of the domain of $\theta_0$ and take $\psi_u:\theta\mapsto \theta(u)$ to be the evaluation functional at $u$ to obtain a simultaneous confidence set for $\theta_0$ over $\mathcal{O}$.

\section{Estimation of the risk functional derivative} \label{sec:derest}

\subsection{Uniform asymptotic linearity of the derivative estimator}

Having outlined a sketch of the proposed framework for inference, we now scrutinize estimation of the G\^{a}teaux derivative of $R_{0}$, which serves as a primary building block of our procedure. To begin, we require that we have at our disposal, for each $h\in\mathcal{H}$, an asymptotically linear estimator $\dot{R}_{n,\theta_*}(h)$ of $\dot{R}_{0,\theta_*}(h)$, in the sense that \begin{align}
\dot{R}_{n,\theta_*}(h) - \dot{R}_{0,\theta_*}(h) = \frac{1}{n} \sum_{i=1}^n \phi_{P_0,\theta_*}(Z_i;h) + r_{n,\theta_*}(h)\ ,
\label{AsympLin}\end{align}where $E_0[ \phi_{P_0,\theta_*}(Z;h)]=0$, $E_0[\phi_{P_0,\theta_*}(Z;h)^2]<\infty$, and $r_{n,\theta_*}(h)=o_P(n^{-1/2})$. The function $z\mapsto \phi_{P_0,\theta_*}(z;h)$ is referred to as the influence function of $\dot{R}_{n,\theta_*}(h)$. In many settings, such an estimator is readily available.
For instance, if the model space $\mathcal{M}$ includes the empirical distribution $P_n$ and the functional $P\mapsto\dot{R}_{P,\theta_*}(h)$ is Hadamard differentiable with respect to the supremum norm \citep[see, e.g.,][]{van2000asymptotic}, the plug-in estimator $\dot{R}_{n,\theta_*}(h) := \dot{R}_{P_n,\theta_*}(h)$ will be asymptotically linear with influence function defined pointwise as \[\phi_{P_0,\theta_*}(z;h)=\left.\frac{d}{d\epsilon}\dot{R}_{P_{\epsilon},\theta_*}(h)\right|_{\epsilon=0} ,\]where $P_\epsilon:=(1-\epsilon)P_0+\epsilon\delta_z$ and $\delta_z$ is a degenerate distribution on $\{z\}$. Hadamard differentiability typically holds in simple examples, such as when the risk functional has the form \eqref{expectation} with loss $\ell_P$ not depending on $P$.
In other problems, the plug-in estimator $\dot{R}_{P_n,\theta_*}(h)$ may fail to even be defined --- this often occurs when the G\^ateaux derivative functional depends on local features of the underlying distribution (e.g., a density or conditional mean function). Provided $P\mapsto \dot{R}_{P,\theta_*}(h)$ is pathwise differentiable relative to the model $\mathcal{M}$ \citep[see, e.g.,][]{bickel1998efficient}, more broadly applicable strategies for estimating $\dot{R}_{0,\theta_*}(h)$ exist. For example, if $\widehat{P}_n\in\mathcal{M}$ is a consistent estimator of $P_0$, possibly obtained via flexible learning strategies (e.g., machine learning), then the one-step debiased estimator \[\dot{R}_{n,\theta_*}(h):=\dot{R}_{\widehat{P}_n,\theta_*}(h)+\frac{1}{n}\sum_{i=1}^{n}\phi_{\widehat{P}_n,\theta_*}(Z_i;h)\] satisfies \eqref{AsympLin} under certain regularity conditions, provided $z\mapsto \phi_{P,\theta_*}(z;h)$ is taken to be any gradient of the pathwise derivative of $P\mapsto \dot{R}_{P,\theta_*}(h)$ \citep{pfanzagl1982contributions}. Alternative constructions with improved properties, such as targeted minimum loss-based estimation (see, e.g., \citealp{van2011targeted}), also exist.

By the central limit theorem, the asymptotic representation \eqref{AsympLin} suffices to establish that, for any finite subset $\mathcal{H}_0\subset \mathcal{H}$,  $\{n^{1/2}[\dot{R}_{n,\theta_*}(h)-\dot{R}_{0,\theta_*}(h)]:h\in\mathcal{H}_0\}$ converges in distribution to a mean-zero Gaussian random vector. This does not readily extend to an infinite set $\mathcal{H}_0$ --- or indeed, $\mathcal{H}$ itself --- without imposing stronger requirements on $\dot{R}_{n,\theta_*}$, and such an extension is needed in our proposal. The following lemma provides additional conditions on $\dot{R}_{n,\theta_*}$ under which this extension holds.

\begin{lemma}\label{weak}
If (i) $\left\{z\mapsto \phi_{0,\theta_*}(z; h): h \in \mathcal{H}\right\}$ is a $P_{0}$-Donsker class, and (ii) $\sup_{h\in\mathcal{H}}|r_n(h)|=o_P(n^{-1/2})$,  then, as an element of $\ell^\infty(\mathcal{H})$, $\{n^{1/2}[\dot{R}_{n,\theta_*}(h) - \dot{R}_{0,\theta_*}(h)]: h \in \mathcal{H}\}$ converges weakly to a tight mean-zero Gaussian process $\mathbb{G}$  with covariance function $\Sigma: (h_1, h_2) \mapsto E_0\,[\phi_{P_0,\theta_*}(Z; h_1)\phi_{P_0,\theta_*}(Z; h_2)]$ relative to the supremum norm.
\end{lemma}
Both conditions constrain the complexity  of $\mathcal{H}$. Condition (i)  often holds, for example, if $\mathcal{H}$ is itself a $P_0$-Donsker class, as implied by Theorem 2.10.6 of \cite{van1996weak}.
Condition (ii)  requires that the asymptotic linearity of $\dot{R}_{n,\theta_*}(h)$ hold uniformly for $h\in\mathcal{H}$. It is trivially satisfied irrespective of $\mathcal{H}$ if, for example, $P_n\in\mathcal{M}$, $P\mapsto \dot{R}_{P,\theta_*}(h)$ is a linear functional, and the plug-in estimator $\dot{R}_{P_n,\theta_*}(h)$ is used.

\subsection{Working examples}

\subsubsection{Example 1: nonparametric mean regression}

We first consider the setting of nonparametric mean regression, in which  $\theta_0$ is the conditional mean function $x\mapsto E_0\,(Y\,|\,X=x)$, which is also expressed as a population risk minimizer in \eqref{RegressionRM}.
For a fixed direction $h$, the G\^ateaux derivative of $\theta\mapsto R_P(\theta)$ at $\theta=\theta_*$ takes the form
$
\dot{R}_{P,\theta_*}(h) = E_P\,\{ [Y - \theta_*(X)] h(X)\}
$.
In particular, this suggests that, in this problem, the score test can be interpreted as examining the orthogonality of the residual calculated under the null hypothesis $H_0:\theta_*=\theta_0$ to all functions $h\in\mathcal{H}$.

The fact that  $P\mapsto \dot{R}_{P,\theta_*}(h)$ is a linear functional defined at the empirical distribution $P_n$ suggests the use of the plug-in estimator 
\begin{align*}
\dot{R}_{n,\theta_*}(h) := \frac{1}{n}\sum_{i=1}^n \left[Y_i - \theta_*(X_i) \right] h(X_i)
\end{align*}
of $\dot{R}_{0,\theta_*}(h)$. This plug-in estimator is in fact unbiased and asymptotically linear with influence function 
\begin{align*}
z=(x,y)\mapsto \phi_{P_0,\theta_*}(z; h):= [y - \theta_*(x)]h(x) - \dot{R}_{0,\theta_*}(h)\ .
\end{align*} Since this influence function is the efficient influence function of $P\mapsto \dot{R}_{P,\theta_*}(h)$ relative to a nonparametric model, $\dot{R}_{n,\theta_*}(h)$ is also nonparametric efficient.
The remainder term $r_n(h)$ in \eqref{AsympLin} is exactly zero, so the conditions of Lemma~\ref{weak} are often satisfied as long as $\mathcal{H}$ satisfies a Donsker condition. Uniform convergence of $\dot{R}_n(\theta_*; h)$ is thus achieved under relatively weak conditions.

%

\subsubsection{Example 2: partially additive mean regression}

We now consider the setting of a partially additive mean model. This example is more involved than the previous example because the risk functional $P\mapsto R_P$ is nonlinear  and depends on $P$ via an unknown function-valued nuisance parameter $f_P$. It is possible to verify that this nuisance parameter can be expressed as $f_P:w\mapsto E_P\,[Y-\theta_P(X)\,|\,W=w]$. Coupled with \eqref{PartAddModRM}, this fact implies that 
$\theta_P$ minimizes the population $P$-risk functional $
R_{P,\theta_*}(h) := E_P\,[Y - \mu_{Y,P}(W) - \{\theta_*(X) - E_P\,[\theta_*(X)\,|\,W] \} ]^2$, where we define $\mu_{Y,P}:w\mapsto E_P\,(Y\,|\,W=w)$.
The G\^ateaux derivative of this risk functional takes the form
\begin{align}
\dot{R}_{P,\theta_*}(h) = E_P\,\{ [Y - \mu_{Y,P}(W)  - \theta_*(X) + \mu_{\theta_*,P}(W) ][h(X) - \mu_{h,P}(W)] \}\ ,
\label{PAM-GD}
\end{align}
where we define $\mu_{g,P}:w\mapsto E_P\,[g(X)\,|\,W=w]$ for each function  $g:\mathbb{R} \rightarrow \mathbb{R}$ for which this moment exists.

Obtaining an estimator of $\dot{R}_{0,\theta_*}(h)$ is more challenging than in the previous example, as we need to estimate the nuisance parameters $\mu_{Y,P_0}$, $\mu_{h,P_0}$ and $\mu_{\theta_*,P_0}$.
Suppose that we have constructed consistent estimators $\mu_{n,Y,P_0}$, $\mu_{n,h,P_0}$ and $\mu_{n,\theta_*,P_0}$ of these nuisance functions using a nonparametric estimation procedure, such as artificial neural networks \citep{barron1989statistical}, the highly adaptive lasso \citep{benkeser2016highly}, or the Super Learner \citep{van2007super}.
The resulting plug-in estimator, 
\begin{align}
\dot{R}_{n,\theta_*}(h) := \frac{1}{n} \sum_{i=1}^n [Y_i - \mu_{n,Y,P_0}(W_i) + \mu_{n,\theta_*,P_0}(W_i)- \theta_*(X_i)]
[h(X_i) - \mu_{n,h,P_0}(W_i)], 
\label{PAM-Plug}
\end{align} 
can be shown to be asymptotically linear with influence function 
\begin{align}
z=(w,x,y)\mapsto \phi_{P_0,\theta_*}(z; h):=\{y - \mu_{Y,P_0}(w) + \mu_{\theta_*,P_0}(w)- \theta_*(x)\}\{h(x) - \mu_{h,P_0}(w) \} - \dot{R}_{0,\theta_*}(h)
\label{PAM-EIF}
\end{align}under rate and complexity conditions on nuisance estimators $\mu_{n,Y,P_0}$, $\mu_{n,h,P_0}$ and $\mu_{n,\theta_*,P_0}$. Again, since this influence function is the efficient influence function $P\mapsto \dot{R}_{P,\theta_*}(h)$ relative to a nonparametric model, $\dot{R}_{n,\theta_*}(h)$ is also nonparametric efficient.

While the asymptotic linearity of $\dot{R}_{n,\theta_*}(h)$ at a fixed $h$ can be readily established, uniform asymptotic linearity over $\mathcal{H}$ is more difficult to achieve and requires stronger conditions, stated explicitly in the Supplementary Material (see Part I).
In particular, uniform control of the remainder from the linear representation of $\dot{R}_{n,\theta_*}(h)$ (condition ii of Lemma 1) requires consistent estimation of the nuisance function $\mu_{h,P_0}$ uniformly over $\mathcal{H}$.
For large $\mathcal{H}$, this may be a difficult feat. 
Additionally, computational difficulties may arise since computing an estimate of $\mu_{h,P_0}$ separately for each $h \in \mathcal{H}$ can be unfeasible unless a clever parametrization of $\mathcal{H}$ is available.
We provide such a construction in Section~\ref{sec:implementation}.

\section{Properties of the restricted score test}\label{sec:properties}

\subsection{Limiting distribution of the test statistic}\label{limit}

We now describe the construction of our restricted score test based on an estimator $\dot{R}_{n,\theta_*}$ satisfying the conditions outlined in Section~\ref{sec:derest}
and establish its large-sample properties.

Whenever $\theta_0=\theta_*$, the test statistic $T_n := \Omega( n^{1/2}\dot{R}_{n,\theta_*})$ converges in distribution to $\Omega\left(\mathbb{G}\right)$ for any norm $\Omega$ on $\ell^\infty(\mathcal{H})$.
We wish to compute the distribution function of $\Omega(\mathbb{G})$ in order to obtain an approximate p-value $\rho(t)$ based on an arbitrary realization $t$ of $T_n$.
However, since the limiting distribution of $T_n$ is generally not available in closed form, we resort instead to resampling techniques.

We propose a multiplier bootstrap procedure that leverages the asymptotic linearity of $\dot{R}_{n,\theta_*}$ to approximate the p-value $\rho(t)$.
For each $m = 1,2,\ldots,M$, let $\xi_{m,1},\xi_{m,2},\ldots,\xi_{m,n}$ be a random sample of independent and identically distributed random variables (also independent of $Z_1,Z_2,\ldots,Z_n$) with mean zero, unit variance and finite moment of order $2+\omega$ for some $\omega>0$. For instance, these could be taken as a random sample of Rademacher or standard normal random variables.
Defining the bootstrapped mapping $\dot{R}_{m,n,\theta_*}:h\mapsto\frac{1}{n}\sum_{i=1}^{n}\phi_{n,\theta_*}(Z_i,h)\xi_{m,i}$, where $\phi_{n, \theta_*}$ a consistent estimator of $\phi_{P_0, \theta_*}$, we construct the bootstrapped test statistic
\begin{align}
T_{m,n} := \Omega\left( n^{1/2}\dot{R}_{m,n,\theta_*}\right)\ .
\label{Bootstrap}
\end{align}
Under suitable regularity conditions, this statistic converges weakly to $\Omega(\mathbb{G})$, and so, 
\begin{align*}
\rho_{M,n}(t):=\frac{1}{M}\sum_{j=1}^M I\left(T_{m,n} > t\right)
\end{align*}
serves as an approximation to $\rho(t)$ for $n$ and $M$ large. 
This result is stated formally in the following theorem. Below, we suppose that the influence function $\phi_{P_0,\theta_*}$ depends on $P_0$ only through some nuisance parameter $f_0\in\mathcal{F}$, where $\mathcal{F}$ is a vector space endowed with some norm $\|\cdot\|_\mathcal{F}$. With some abuse of notation, for any candidate nuisance $f\in\mathcal{F}$, we denote by $\phi_{f,\theta_*}$ the influence function corresponding to nuisance value $f$. We note then, in particular, that $\phi_{f_0,\theta_*}=\phi_{P_0,\theta_*}$. We consider this representation of $\phi_{P_0,\theta_*}$ to leverage the fact that, in practice, to implement our procedure, we may not need to estimate the entire distribution $P_0$ but rather some summary of $P_0$ (e.g., a mean value or conditional mean function under $P_0$). This does not put any additional restriction on the problem, though, as we may also take $f_0$ to be the density function of $P_0$, if appropriate. We suppose that we have access to an estimator $f_n$ of $f_0$ based on $Z_1,Z_2,\ldots,Z_n$.

\begin{theorem}
Let $\xi_1,\xi_2,\ldots,\xi_n$ be independent and identically distributed random variables  with mean zero, variance one and finite raw moment of order $2+\omega$ for some $\omega > 0$, and also independent of $Z_1,Z_2,\ldots,Z_n$. Suppose that, for some $\delta>0$, the class $
\Phi_{\delta} := \left\{z\mapsto\phi_{f,\theta_*}(z; h) - \phi_{f_0,\theta_*}(z; h): h \in \mathcal{H}, f \in \mathcal{F},\|f - f_{0} \|_{\mathcal{F}} < \delta \right\}$
is $P_0$-Donsker and has a finite envelope function. 
Then, provided that  $\|f_{n} - f_{0} \|_{\mathcal{F}} = o_P(1)$ and that
\begin{align*}
\sup_{h \in \mathcal{H}} \int[\phi_{f,\theta_*}(z;h) - \phi_{f_0,\theta_*}(z;h)]^2 dP_0(z) \longrightarrow 0
\end{align*}
as $\|f-f_0\|_\mathcal{F}\rightarrow 0$,
$\left\{n^{-1/2} \sum_{i=1}^n \xi_i \phi_{f_n,\theta_*}\left(Z_i; h\right): h \in \mathcal{H} \right\}$ converges weakly to $\mathbb{G}$ relative to the supremum norm as an element of $\ell^\infty(\mathcal{H})$ conditional upon the sample paths $Z_1,Z_2,\ldots,Z_n$, in outer probability.
\end{theorem}

We recall that the Gaussian process $\mathbb{G}$ was explicitly defined in Lemma 1. In view of this result, the resampling-based strategy described above can be used to approximate the limiting distribution of the test statistic under $H_0:\theta_0=\theta_*$. However, given that  the limiting distributions of $\Omega(n^{1/2}\dot{R}_{n,\theta_*})$ and $\Omega(n^{1/2}\dot{R}_{n,\theta_0})$  coincide under the null hypothesis, we could  instead obtain and use an approximation to the latter. This strategy has the key advantage that the same bootstrap samples can be used to test $H_0: \theta_0 = \theta_*$ for \textit{any} value of $\theta_*$.
This can lead to large gains in  computational efficiency when constructing confidence bands, as we discuss in Section \ref{confidence}.
Using the tools used to prove Theorem 1, it is possible to derive a slight modification of the result in which $\theta_*$ is replaced by a consistent estimator $\theta_n$.

\subsection{Selection of the class of directions and norm}

While our proposed test of $H_{0,r}$ --- and thus of $H_0$ --- achieves nominal type I error control with any norm $\Omega$ and any sufficiently small class $\mathcal{H}$, statistical power is influenced by these selections.
The optimal norm and class of directions as well as the sensitivity of power to their selection depend on properties of the underlying data-generating mechanism $P_0$. 


We first discuss selection of $\mathcal{H}$. To do so, we examine the local asymptotic power of the restricted score test when $\mathcal{H}$ is a singleton set containing a fixed direction $h$.
Suppose that $Z_1,Z_2,\ldots, Z_n$ are generated from a distribution $P_{0,n}$ such that the G\^ateaux derivative under $P_{0,n}$ is $n^{-1/2}t_h$ for $t_h \in \mathbb{R}$.
Suppose also that $\dot{R}_{n, \theta_*}(h)$ is locally regular in the sense that \begin{align*}
    \dot{R}_{n, \theta_*}(h) - n^{-1/2}t_h = \frac{1}{n}\sum_{i=1}^n \phi_{0, \theta_0}(Z_i;h) + r_n(h)\ ,
\end{align*}
where the influence function $\phi_{0, \theta_0}$ does not depend on $n$, and $n^{1/2}r_n(h)$ converges to zero in probability under sampling from $P_{0,n}$.
It can be shown that the most powerful test of the null hypothesis that $\dot{R}_{0,\theta_*}(h) = 0$ rejects the null when the statistic $T_n^*:= \dot{R}^2_{n,\theta_*}(h)/E_0\,[\phi^2_{0,\theta_0}(Z; h)]$ is larger than the $(1-\alpha)$-quantile of the $\chi^2_1$ distribution, and that $T_n^*$ approximately follows a non-central chi-squared distribution with one degree of freedom and non-centrality parameter $t^2_h/E_0\,[\phi^2_{0,\theta_0}(Z; h)]$ for large $n$.
It can thus be seen that the local asymptotic power is determined by the ratio of the G\^ateaux derivative to the asymptotic variance of $n^{1/2}[\dot{R}_{n,\theta_0}(h)-\dot{R}_{0,\theta_0}(h)]$,
\begin{align}
 \frac{\dot{R}_{0,\theta_*}^2(h)}{E_0\,[\phi^2_{0,\theta_0}(Z; h)]}
 \label{MaxGateauxDeriv}\ .
\end{align}Thus, if we perform a test by taking $\mathcal{H}$ to be a set containing only a single fixed direction, an optimal direction would be any maximizer $h_0$ of \eqref{MaxGateauxDeriv} over $h \in \mathcal{H}(\theta_*)$.
It is therefore reasonable to seek to select $\mathcal{H}$ as a small set of functions that contains a good approximation of $h_0$.
Because $h_0$ is unknown and will typically depend on $P_0$, a possible approach consists of characterizing $h_0$ analytically and then taking $\mathcal{H}$ to be a class of smooth functions that contains an estimate of $h_0$.  We provide explicit details regarding the construction of $\mathcal{H}$ in Section 5.

We now provide natural examples of the norm $\Omega$. Let $V:\mathcal{H} \to [0,\infty)$ be a non-negative weight functional.
We first consider the weighted supremum norm
\begin{align*}
a\mapsto \Omega_{\infty}(a) := \sup_{h \in \mathcal{H}} \, V(h)|a(h)|\ ,
\end{align*}
leading to the test statistic $\Omega_{\infty}(n^{1/2}\dot{R}_{n,\theta_*}) = n^{1/2}\sup_{h \in \mathcal{H}} V(h) |\dot{R}_{n,\theta_*}( h)|$.
With the choice $V\equiv 1$, this simply evaluates $n^{1/2}\dot{R}_{n,\theta_*}$ at the direction of greatest estimated change.
Under the alternative hypothesis, the largest estimated G\^ateaux derivative value does not necessarily provide the greatest evidence in favor of the alternative because of the variability of the derivative estimator. This motivates us to instead weight the G\^ateaux derivative by the reciprocal of the standard deviation implied by the influence function of the derivative estimator, thus setting $V=V_0$ with $V_0(h) := \{E_0\,[\phi_{P_0, \theta_0}(Z;h)^2]\}^{-1/2}$.
We note that, when $\Theta$ is finite-dimensional and $\mathcal{H} = \mathcal{H}(\theta_*)$, the statistic resulting from use of the variance-weighted supremum norm is precisely equivalent to the original score statistic proposed in \cite{rao1948large}.
The variance-weighted supremum norm can thus be viewed as a natural generalization of the standard parametric score test to the infinite-dimensional setting.
We note that our suggested choice of weight $V_0$ depends on $P_0$, and so, in practice, we must use an estimator $V_n$ of $V_0$.
It can be seen through an application of Slutsky's theorem that, as long as $V_n$ is uniformly consistent, in the sense that $\sup_{h \in \mathcal{H}} \left|V_n(h) - V_0(h)\right| = o_P(1)$, our theoretical results remain valid. 

As an alternative norm, we also consider a weighted $L_2$ norm over $\mathcal{H}$. Let $Q$ be a measure on the Borel $\sigma$-algebra generated by $\mathcal{H}$.
We define the weighted $L_2$ norm as
\begin{align*}
a\mapsto\Omega_2(a) := \left\{\int_{\mathcal{H}} \, [V(h)a(h)]^2 dQ(h)\right\}^{1/2}
\end{align*} and consider the test statistic $\Omega_2(n^{1/2}\dot{R}_{n,\theta_*})$.
This statistic involves weighted averaging of the G\^ateaux derivative corresponding to each direction $h \in \mathcal{H}$.
Similarly as with the supremum norm, we may wish to set $V=V_0$ in order to place more weight on directions for which we can estimate the G\^ateaux derivative with greater precision.

To understand the influence of the choice of $\Omega$ on power, we draw intuition from literature on simultaneous testing in the high-dimensional setting \citep[e.g.,][]{cai2014two}.
In settings where the signal is dense, in the sense that the G\^ateaux derivative is small but nonzero in many directions in $\mathcal{H}$, good performance is expected from the $L_2$ norm but not from the supremum norm. Conversely, when the signal is sparse, in the sense that the G\^ateaux derivative is large in relatively few directions and zero elsewhere, the supremum norm is expected to yield better performance than the $L_2$ norm.

\subsection{Extension to data-dependent classes of directions}\label{randomsets}

So far, we have considered $\mathcal{H}$ to be a fixed class. In practice, it can be difficult to select $\mathcal{H}$ a priori, and we may want to instead select the class of directions in a data-driven manner. To be applicable in such cases, our theoretical results must allow the fixed class $\mathcal{H}$ to be replaced by a stochastic (data-dependent) sequence of classes $\mathcal{H}_{n}=\mathcal{H}_n(Z_1,Z_2,\ldots,Z_n)$. The following theorem indicates that if $\mathcal{H}_n$ converges to a fixed class $\mathcal{H}$ in an appropriate sense, then for the two choices of norm we have considered, $\Omega(n^{1/2}\dot{R}_{n,\theta_*})$ also converges weakly to $\Omega\left(\mathbb{G}\right)$, where $\mathbb{G}$ is the same Gaussian process defined in Lemma \ref{weak}. Below, to simplify the notation, we fix $\theta_*$ and denote by $\phi_h$ the function $z\mapsto \phi_{P_0,\theta_*}(z;h)$.

\begin{theorem}
Suppose that there exists a function class $\bar{\mathcal{H}}$ such that $\{\phi_h: h \in \bar{\mathcal{H}} \}$ is a $P_0$-Donsker class with a finite and square-integrable envelope, and that $\mathcal{H}_n \cup \mathcal{H} \subseteq \bar{\mathcal{H}}$ with $P_0$-probability one.  Suppose also that $H_0:\theta_*=\theta_0$ holds.
\begin{enumerate}[(a)]
\item If $\{\phi_{h}: h \in \mathcal{H}_{n}\}$ converges to $\{\phi_{h}: h \in \mathcal{H}\}$ in the Hausdorff sense, that is,
\begin{align}
\max\Bigg\{ &\sup_{h_1 \in \mathcal{H}} \inf_{h_2 \in \mathcal{H}_{n}} \int [\phi_{h_1}(z)-\phi_{h_2}(z)]^2dP_0(z),
\sup_{h_2 \in \mathcal{H}_n} \inf_{h_1 \in \mathcal{H}}\int [\phi_{h_1}(z)-\phi_{h_2}(z)]^2dP_0(z)
 \Bigg\} = o_P(1)\ ,
 \label{hausdorff}
\end{align}then $\sup_{h \in \mathcal{H}_n} n^{1/2} \int \phi_{h}(z) d(P_n - P_0)(z)$ converges in distribution to $\sup_{h \in \mathcal{H}} \mathbb{G}(\phi_h)$.
\item Let $\mathcal{B}(\bar{\mathcal{H}})$ denote the Borel $\sigma$-algebra, and let $\bar{Q}$ be a measure on $\mathcal{B}(\bar{\mathcal{H}})$. If \begin{align}
 \bar{Q}\left(\left\{\mathcal{H}\cup\mathcal{H}_n\right\}\setminus\left\{\mathcal{H}\cap\mathcal{H}_n\right\}\right) = o_P(1)\ ,
 \label{QbarDist}
\end{align}
then $\int_{\mathcal{H}_n}  n\{\int \phi_{h}(z)d(P_n - P_0)(z)\}^2 d\bar{Q}(h)$ converges in distribution to $\int_\mathcal{H} \int\{\mathbb{G}(\phi_h) \}^2 d\bar{Q}(h)$. 
\end{enumerate}\label{thm:adaptive}
\end{theorem}
Theorem 2 can be applied to conclude that $\Omega\left(n^{1/2}\dot{R}_{n,\theta_*}\right)$ converges weakly to $\Omega\left(\mathbb{G}\right)$ if $\dot{R}_{n,\theta_*}(h)$ is uniformly asymptotically linear for $h \in \bar{\mathcal{H}}$, with $\bar{\mathcal{H}}$ defined in the theorem statement.
For convergence of the supremum norm, Theorem~\ref{thm:adaptive} requires that for any direction $h_1 \in \mathcal{H}$, there exists a direction $h_2 \in \mathcal{H}_n$ such that the expected squared difference between the influence functions for the G\^ateaux derivative estimator corresponding to directions $h_1$ and $h_2$ converges in probability to zero, and similarly for any direction $h_2 \in \mathcal{H}_n$.
For convergence of the $L_2$ norm, we require that the measure of the difference between the union and intersection of $\mathcal{H}$ and $\mathcal{H}_n$ converges in probability to zero.
We expect that, under suitable regularity conditions, the multiplier bootstrap scheme described in Section \ref{limit} also provides a valid approximation of the sampling distribution of $\Omega(n^{1/2}\dot{R}_{n,\theta_*})$ when the class of directions is data-dependent.

\subsection{Construction of confidence regions}\label{confidence}

By taking advantage of the relationship between hypothesis tests and confidence regions, we can invert the proposed score test to obtain simultaneous confidence sets for summaries of $\theta_0$.
Let $\Psi:= \{\psi_u: u \in \mathcal{U} \}$ denote a collection of real-valued functionals defined on $\Theta$ indexed by some set $\mathcal{U}$. We wish to construct simultaneous confidence intervals for elements of $\Psi(\theta_0):=\{\psi_u(\theta_0):u\in\mathcal{U}\}$.

As before, we take $\mathcal{C}_{n}:=\{\theta\in\Theta:\text{we fail to reject $\theta_0=\theta$ against $\theta_0\neq \theta$ based on $Z_1,Z_2,\ldots,Z_n$}\}$ to denote the set of null parameter values that the restricted score test fails to reject.
If the test achieves the nominal type I error rate $\alpha$, $\theta_0$ belongs to $\mathcal{C}_n$ with probability tending to $1-\alpha$ in the sense that
\begin{align*}
P_0\,\big(\theta_0 \in \mathcal{C}_n) = P_0\,(\text{we fail to reject $\theta_0=\theta$ against $\theta_0\neq \theta$ based on $Z_1,Z_2,\ldots,Z_n$}\big) \longrightarrow 1-\alpha
\end{align*}as sample size $n$ tends to infinity whenever $H_0:\theta_0=\theta$ is true.
Thus, $\mathcal{C}_n$ is a $100 (1-\alpha)\%$ confidence region for $\theta_0$.
The random region $\mathcal{C}_n$ can be interpreted as the collection of parameter values $\theta$ that are consistent with the observed data. To obtain a confidence region for $\Psi(\theta_0)$, for each $u\in\mathcal{U}$, we find the largest and smallest value of $\psi_u(\theta)$ that can be obtained for $\theta \in \mathcal{C}_n$; setting $\Psi_{n,u}:=\{\psi_u(\theta):\theta\in\mathcal{C}_n\}$, we construct the set 
\begin{align*}
\mathcal{C}_n(u) := 
(
\inf\Psi_{n,u},\ \sup\Psi_{n,u}
)\ .
\end{align*}
To see that $\{\mathcal{C}_n(u): u \in \mathcal{U}\}$ is in fact a simultaneous confidence region, we note that if  $\psi_{u'}(\theta_0) \notin \mathcal{C}_n(u')$ for some $u'\in\mathcal{U}$, then it is necessarily the case that $\theta_0 \notin \mathcal{C}_n$.
Thus, we have that
\begin{align*}
\liminf_{n\to\infty}P_0\,\{ \psi_u(\theta_0) \in \mathcal{C}_n(u) \text{ for all } u \in \mathcal{U}\}\ \geq\  \liminf_{n\to\infty}P_0\,(\theta_0 \in \mathcal{C}_n)\ =\ 1-\alpha\ .
\end{align*}

In some instances, if $\Theta$ is unrestricted, an interval $\mathcal{C}_n(u)$ can be infinitely wide.
It may be possible to construct a function $\theta$ such that $\dot{R}_{n,\theta}(h) = 0$ for all $h$.
For instance, in Example 1,  this can be achieved by fixing $\theta(X_i) = Y_i$ for $i = 1,2,\ldots,n$, and letting $\theta$ take \textit{any} value elsewhere.
Thus, one can select $\theta \in \mathcal{C}_n$ such that the evaluation $\theta(x_0)$ of $\theta$ at a point $x_0$ where no data are observed is arbitrarily large or small.
This difficulty is avoided if $\Theta$ is a class of smooth functions, although determining such a class \textit{a priori} may be difficult in practice.
If we could obtain a consistent estimator of a measure of the smoothness of $\theta_0$, we could construct a confidence band for a class containing functions no smoother than the smoothness level prescribed by our estimate.
However, it can be challenging to consistently estimate the smoothness of an unknown function.
In our implementation, we use a simple plug-in estimator of the smoothness and leave the development of a more rigorous approach as future work. 

Our proposal is particularly useful for obtaining simultaneous confidence intervals for the evaluation functional $\theta\mapsto \theta(x)$ for arbitrary $x$, as we are able to immediately obtain a $100 (1-\alpha) \%$ confidence band for $\theta_0$.
When only a finite collection of functionals is of interest, however, our proposed intervals may exhibit over-coverage. If the functionals of interest are pathwise differentiable, we recommend instead estimating each quantity using an efficient estimator (as described in, e.g., \citealp{pfanzagl1982contributions} and \citealp{van2011targeted}) to obtain asymptotically calibrated  intervals.

\section{Implementation and practical considerations}\label{sec:implementation}

We now describe our implementation of the restricted score test and discuss its use in the partially additive mean regression example.

\subsection{Construction of $\mathcal{H}$}

For a positive semidefinite kernel function $K$, let $\mathcal{S}_K$ denote its unique reproducing kernel Hilbert space (RKHS), endowed with the inner product $\langle \cdot, \cdot \rangle_{\mathcal{S}_K}$.
We construct $\mathcal{H}$ to be a subspace of $\mathcal{S}_K$.

We consider the eigen-decomposition of $K$ given by
\begin{align*}
(z_1,z_2)\mapsto K(z_1, z_2) = \sum_{j=1}^\infty \kappa_j \eta_j(z_1) \eta_j(z_2)
\end{align*}
with eigenfunctions $\{\eta_1,\eta_2,\ldots\}$ orthogonal with respect to $\langle \cdot, \cdot \rangle_{\mathcal{S}_K}$ and eigenvalues $0\leq \kappa_1<\kappa_2<\ldots$.
Each function $g \in \mathcal{S}_K$ can be expressed as a linear combination   $z\mapsto\sum_{j=1}^\infty a_j \eta_j(z)$ of eigenfunctions, where $a_1,a_2,\ldots$ are real-valued coefficients.
The smoothness of $g$ can be measured by the RKHS norm as
\begin{align*}
J(g) := \langle g, g \rangle_{\mathcal{S}_K} = \sum_{j=1}^\infty\frac{a_j^2}{\kappa_j}
\end{align*}
with higher values of $J(g)$ corresponding to lesser smoothness. The test statistics we consider are based on the ratio of the G\^ateaux derivative estimates to their standard errors and do not depend on the scale of the direction function $h$. Rather, the performance of our test is determined by the \textit{shape} of the direction function.  We therefore require a scale-free measure of smoothness, for which we use a scaled version of the RKHS norm $J(h)V^2_0(h)$, where we recall that  $V_0(h) := \{E_0\,[\phi_{P_0, \theta_0}(Z;h)^2]\}^{-1/2}$ is the reciprocal of the asymptotic standard deviation of $n^{1/2}[\dot{R}_{n,\theta_0}(h) - \dot{R}_{0,\theta}(h)]$.  While other scale-free measurements of smoothness could alternatively be used, we will see that this particular choice leads to computational benefits.
We consider as $\mathcal{H}$ a subset of functions in $\mathcal{S}_K$ with bounded smoothness, namely
\begin{align*}
\mathcal{H}_{\gamma} := \left\{h = \textstyle\sum_{j=1}^\infty a_j \eta_j: a_1,a_2,\ldots\in\mathbb{R},J(h)V^2_0(h) \leq \gamma\right\}
\end{align*}
for some tuning value $\gamma > 0$.
For computational ease, we truncate the eigenbasis at some large level $d$.

In some instances, the kernel for an RKHS has eigenfunctions that are known and available in closed form \citep{wahba1990spline}.
For instance, consider the second-order Sobolev space on $[0,1]$, which can be defined as an RKHS endowed with the inner product $(h_1,h_2)\mapsto\langle h_1, h_2 \rangle_{\mathcal{S}_K} = \int_{0}^1 \ddot{h}_1(z) \ddot{h}_2(z) dz$, where $\ddot{h}$ denotes the second derivative of any given function $h$.
The eigenfunctions and eigenvalues for the kernel are
\begin{align*}
\eta_{2j - 1}:z\mapsto\sqrt{2}\cos\left(2 \pi j z \right) ,\quad \eta_{2j}:z\mapsto\sqrt{2} \sin(2 \pi j z), \quad
\kappa_{2j - 1} = \kappa_{2j} = (2\pi j)^{-4},
\end{align*}for $j=1,2,\ldots$.
When the eigenfunctions are not available in closed form, we can instead use an approximation.

For the partially additive mean model, estimation of the G\^ateaux derivative requires estimation of the conditional mean $\mu_{h,P_0}$ of $h(X)$ given $W$ under $P_0$.
We proceed by first obtaining an estimate $\mu_{n,\eta_j,P_0}$ of the eigenfunction regression function $w\mapsto \mu_{\eta_j, P_0}(w):=E_{0}\left[\eta_j(X)\,|\,W=w\right]$ and then setting $\mu_{n,\eta,P_0} := \sum_{j=1}^d a_j \mu_{n,\eta_j,P_0}$.
The estimate $\mu_{n,\eta_j,P_0}$ may be unreliable for a non-smooth eigenfunction $\eta_j$ and hence lead to a poor estimate of $\mu_{h,P_0}$.
However, by requiring that $h$ is smooth, and thus forcing $a_j$ to be small for non-smooth eigenfunctions, estimates of the non-smooth eigenfunctions should make a relatively small contribution to ${\mu}_{n,h,P_0}$.


We conclude by discussing selection of the tuning parameter $\gamma$.
We choose $\gamma$ so that $\mathcal{H}_{\gamma}$ contains an approximation of a maximizer $h_0$ of \eqref{MaxGateauxDeriv}.
As $h_0$ may depend on $P_0$, we instead obtain an estimate $h_n$ of $h_0$ and take $\gamma = \gamma_n := J\left(h_n\right)V^2_n(h_n)$, where $V_n(h)$ is an estimate of $V_0(h)$.
It can be shown, by an application of the Cauchy-Schwarz inequality,  that for the partially additive mean model, if the residual $Y - f_0(W) - \theta_0(X)$ depends neither on $X$ nor $W$, any maximizer of \eqref{MaxGateauxDeriv} is proportional to $\theta_0 - \theta_*$.
We then set $h_0 = \theta_0 - \theta_*$ and use a simple penalization approach to estimate $\theta_0$, taking
\begin{align}
&(\hat{a}_1,\hat{a}_2,\ldots,\hat{a}_d) := \underset{(a_1,a_2,\ldots,a_d) \in \mathbb{R}}{\argmin}\, R_{n}\left(\textstyle\sum_{j=1}^d a_j \eta_j\right) + \lambda \sum_{j=1}^d \frac{a^2_j}{\kappa_j} \label{PenOpt}
\end{align}and setting $\theta_n := \sum_{j=1}^d \hat{a}_j \eta_j$,
where $R_n$ is an estimate of the population risk function $R_0$, and the tunning parameter $\lambda > 0$ is chosen by cross-validation. 
We then take $h_n := \theta_n - \theta_*$.

%
%


\subsection{Calculation of the test statistic}\label{sec:teststatcal}

We now discuss strategies for computing the test statistic.
As calculation of the test statistic depends on the form of $\dot{R}_{n,\theta_*}(h)$, we focus specifically on the partially additive mean model.

The estimate $\dot{R}_{n,\theta_*}(h)$ of the G\^ateaux derivative \eqref{PAM-GD} can be expressed as
\begin{align*}
\dot{R}_{n,\theta_*}(h) =  n^{-1}\mathbf{S}(\theta_*)^\top \Gamma \mathbf{a}
\end{align*}
for each $h \in \mathcal{H}_\gamma$, where $\mathbf{S}(\theta_*)$ is an $n$-dimensional vector with $i^{th}$ component  $\mathbf{S}(\theta_*)_i:= Y_i - \mu_{n,Y,P_0}(W_i) - [\theta_*(X_i) - \mu_{n,\theta_*,P_0}(X_i)]$, $\Gamma$ is an $n \times d$ matrix with $(i,j)^{th}$ entry $\Gamma_{ij} := \eta_{j}(X_i) - \mu_{n,\eta_j,P_0}(W_i)$, and $\mathbf{a}$ is a $d$-dimensional vector of coefficients for the eigenbasis.
We estimate the variance of the efficient influence function $\phi_{P_0, \theta_0}(Z; h)$ as 
\begin{align*}
    V^{-2}_n(h) = \frac{1}{n}\sum_{i=1}^n \left\{[Y_i - \mu_{n,Y,P_0}(W_i) + \mu_{n,\theta_n,P_0}(W_i)- \theta_n(X_i)]
[h(X_i) - \mu_{n,h,P_0}(W_i)]\right\}^2.
\end{align*}
For $h \in \mathcal{H}_\gamma$, we can rewrite the estimate as $V_n^{-2}(h) = \mathbf{a}^\top \mathbf{V} \mathbf{a}$, where $\mathbf{V} = n^{-1}\Gamma^\top[\text{diag}\left(\mathbf{S}(\theta_n)\right)]^2\Gamma$ .

The inverse variance-weighted supremum norm test statistic $\omega_{\infty,n}:=\Omega_{\infty}(n^{1/2}\dot{R}_{n,\theta_*})$ then takes the form
\begin{align}
\omega^2_{\infty,n} = \sup_{\mathbf{a}} \left\{ \frac{n^{-1}[\mathbf{S}(\theta_*)^\top \Gamma \mathbf{a}]^2}{\mathbf{a}^\top \mathbf{V} \mathbf{a}}: \frac{\mathbf{a}^\top\text{diag}\left(\frac{1}{\boldsymbol{\kappa}}\right)\mathbf{a}}{\mathbf{a}^\top \mathbf{V} \mathbf{a}} \leq \gamma \right\},
\label{WeightSupNormLinear}
\end{align}
where $\boldsymbol{\kappa} := (\kappa_1,\kappa_2,\ldots,\kappa_d)$.
A maximizer of the optimization problem in \eqref{WeightSupNormLinear} can be obtained by solving
\begin{align}
\sup_{\mathbf{a}}\left\{ n^{-1/2}\mathbf{S}(\theta_*)^\top\Gamma \mathbf{a}: \mathbf{a}^\top \mathbf{V} \mathbf{a} = 1, \mathbf{a}^\top\text{diag}\left(\frac{1}{\boldsymbol{\kappa}}\right)\mathbf{a} \leq \gamma \right\}.
\label{ApproxSupNorm1}
\end{align}
%
%
By writing the constrained optimization problem \eqref{ApproxSupNorm1} in the Lagrangian form, it can be shown that the maximizer of \eqref{ApproxSupNorm1} is the solution to
\begin{align}
\underset{\mathbf{a}}{\argmax}\left\{
n^{-1/2}\mathbf{S}(\theta_*)^\top \Gamma \mathbf{a} - \frac{\lambda_2}{2}\,\left( \mathbf{a}^\top \mathbf{V} \mathbf{a}  + \lambda_1\, \mathbf{a}^\top\text{diag}\left(\frac{1}{\boldsymbol{\kappa}}\right)\mathbf{a}\right)\right\},
\label{PenSupNorm}
\end{align}
where $\lambda_1,\lambda_2 > 0$ are chosen so that the constraints in \eqref{ApproxSupNorm1} are satisfied.
With some algebra, we find that the solution $\tilde{\mathbf{a}}_{\lambda_1, \lambda_2}$ to \eqref{PenSupNorm} is available in closed form as
\begin{align*}
\tilde{\mathbf{a}}_{\lambda_1, \lambda_2} = n^{-1/2}\lambda_2^{-1}\left\{\,\mathbf{V} + \lambda_1\,\text{diag}\left(\frac{1}{\boldsymbol{\kappa}}\right)  \right\}^{-1} \Gamma^\top \mathbf{S}(\theta_*)\ .
\end{align*}
Thus, the supremum norm test statistic can be expressed as
\begin{align}
\omega_{\infty,n} = n^{-1}\lambda_2^{-1}\mathbf{S}(\theta_*)^\top \Gamma \left\{\,\mathbf{V} + \lambda_1\,\text{diag}\left(\frac{1}{\boldsymbol{\kappa}}\right)  \right\}^{-1} \Gamma^\top \mathbf{S}(\theta_*)\ .
\label{QuadFormSupNorm}
\end{align}
If we fix $\lambda_1$ and $\lambda_2$, we obtain a test statistic that can be expressed as a quadratic form in $\mathbf{S}(\theta_*)$.
Fixing $\lambda_1$ and $\lambda_2$ is appealing due to computational difficulties with confidence band construction arising when the test statistic is not available in closed form, as discussed in Section 5.4.

One can see in \eqref{QuadFormSupNorm} that $\lambda_2$ affects the scale of $\omega_{\infty, n}$ but has no other effect on its sampling distribution.
Thus, there is only a single relevant tuning parameter $\lambda_1$, which governs the trade-off between the RKHS norm $J(h)$ and the variance of the G\^ateaux derivative estimator $V^{-2}_0(h)$.
In fact, with $\lambda_1$ fixed, the test statistic is proportional to
\begin{align*}
\sup_{\mathbf{a}} \left\{ \frac{n^{-1}[\mathbf{S}(\theta_*)^\top (\Gamma \mathbf{a})]^2}{\mathbf{a}^\top \mathbf{V} \mathbf{a} + \lambda_1 \mathbf{a}^\top \text{diag}\left(\frac{1}{\boldsymbol{\kappa}}\right) \mathbf{a}} \right\}
\end{align*}
and can thus be viewed as a penalized version of the original supremum norm test statistic in \eqref{WeightSupNormLinear}.
In our implementation, we use this penalized supremum norm test statistic and take $\lambda_1$ as the solution to
\begin{align*}
    \frac{\tilde{\mathbf{a}}^\top_{\lambda_1, \lambda_2} \text{diag}\left(\frac{1}{\boldsymbol{\kappa}}\right)\tilde{\mathbf{a}}_{\lambda_1, \lambda_2}}{\tilde{\mathbf{a}}_{\lambda_1, \lambda_2}^\top\mathbf{V} \tilde{\mathbf{a}}_{\lambda_1, \lambda_2}} = \gamma.
\end{align*}
Though this choice is data-adaptive, it is theoretically justified as long as this data-adaptive choice of $\lambda_1$ converges in probability to a constant.

To approximate the $L_2$ norm test statistic $\Omega_2(n^{1/2}\dot{R}_{n, \theta_*})$, we use a Monte Carlo sampling algorithm.
Let $A$ be a $N(0, \{\mathbf{V} + \lambda_3\,\text{diag}\left(\frac{1}{\boldsymbol{\kappa}}\right)\}^{-1} )$ random variable, where $\lambda_3 > 0$ is a tuning parameter that modulates the distribution of $\frac{A^\top \text{diag}\left(\frac{1}{\boldsymbol{\kappa}}\right) A}{A^\top \mathbf{V} A}$, and let $\mathbf{a}_1,\ldots,\mathbf{a}_B$ be a sample of independent draws from the distribution of $A$.
For large $\lambda_3$, $\frac{\mathbf{a}^\top_b \text{diag}\left(\frac{1}{\boldsymbol{\kappa}}\right) \mathbf{a}_b}{\mathbf{a}_b^\top \mathbf{V} \mathbf{a}_b}$ will be small for a large proportion of the sample.
We select $\lambda_3$ so that a large portion of the vectors $\mathbf{a}_b$, say half, belongs to the set $\mathcal{A}_\gamma = \left\{\mathbf{a}: \mathbf{a}^\top \text{diag}\left(\frac{1}{\boldsymbol{\kappa}}\right) \mathbf{a} \leq \gamma \mathbf{a}^\top \mathbf{V} \mathbf{a}\right\}$.
Let $\pi$ be an approximation of the density function of $\frac{A^\top \text{diag}\left(\frac{1}{\boldsymbol{\kappa}}\right) A}{A^\top \mathbf{V} A}$, and let $\pi_b  = \pi\left(\frac{\mathbf{a}_b^\top \text{diag}\left(\frac{1}{\boldsymbol{\kappa}}\right) \mathbf{a}_b}{\mathbf{a}_b^\top \mathbf{V} \mathbf{a}_b}\right)$.
We consider the approximated $L_2$ norm test statistic 
\begin{align}
\omega_{2,n,B}:= \sum_{b=1}^B
\frac{[\mathbf{S}^\top(\theta_*) \Gamma \mathbf{a}_b]^2}{ \mathbf{a}_b^\top \mathbf{V} \mathbf{a}_b}\frac{I\left(\mathbf{a}_b \in \mathcal{A}_{\gamma}\right)}{\pi_b}\ .
\label{ApproxL2Norm}
\end{align}
The $L_2$ norm $\omega_{2,n,B}$ is an average of squared G\^ateaux derivative estimates for a random sample of directions $h \in \mathcal{H}_\gamma$, inversely weighted by the variance of the estimates and the density of $J(h)V^2(h)$.
Inversely weighting by the density of $J(h)V^2(h)$ results in directions with equal smoothness receiving equal weight and reduces the impact of $\lambda_3$ on the sampling distribution of the test statistic.
The approximated $L_2$ norm test statistic is also available in quadratic form in $\mathbf{S}(\theta_*)$ as $\omega_{2,n,B} = n^{-1}\left(\Gamma^\top\mathbf{S}(\theta_*)\right)^\top  \mathbf{A}^\top\mathbf{U}^{-1} \mathbf{A}  \left(\Gamma^\top\mathbf{S}(\theta_*)\right)$,
where $\mathbf{A}$ is a $B\times d$ matrix with $(b,j)^{th}$ element $\mathbf{a}_{b,j}$ and $\mathbf{U}$ is a $B$-dimensional diagonal matrix  with $b^{th}$ diagonal entry $\mathbf{U}_{b} := \pi_b\mathbf{a}_b^\top \mathbf{V} \mathbf{a}_b$.

The distribution of $\mathbf{a}_b$ can influence the statistical power of the test, and alternative approaches for generating the Monte Carlo sample can be considered.
For example, if we have prior knowledge about which directions provide strong evidence in favor of the alternative hypothesis, we may choose to generate Monte Carlo samples from a distribution that places more weight on such directions.  When relevant prior knowledge is not available, one may prefer to use the supremum norm.

%

\subsection{Calculation of the multiplier bootstrap test statistics}

The bootstrap test statistic \eqref{Bootstrap} can be computed using a similar strategy as in Section~\ref{sec:teststatcal}.
Let $\xi_1,\xi_2,\ldots,\xi_n$ be a sample of independent draws from the standard normal distribution.
The multiplier bootstrap derivative estimate is
\begin{align*}
  \dot{R}_{m,n,\theta_n}(h) = \frac{1}{n}\sum_{i=1}^n \xi_i\left\{[Y_i - \mu_{n,Y,P_0}(W_i) + \mu_{n,\theta_n,P_0}(W_i)- \theta_n(X_i)]
[h(X_i) - \mu_{n,h,P_0}(W_i)] - \dot{R}_{n,\theta_n}(h)\right\}.
\end{align*}
For $h \in \mathcal{H}_\gamma$, we can re-write $ \dot{R}_{m,n,\theta_n}(h)$ as $\mathbf{S}(\theta_n)^\top  \mathrm{diag}\left(\boldsymbol{\xi} - \bar{\xi}\,\right) \Gamma \mathbf{a}$, where $\boldsymbol{\xi}:= (\xi_1,\xi_2,\ldots,\xi_n)$, $\bar{\xi} := \frac{1}{n}\sum_{i=1}^n \xi_i$, and $\mathbf{S}$, $\Gamma$ and $\mathbf{a}$ are as defined in Section~\ref{sec:teststatcal}.
Thus, the bootstrap test statistics can be computed using the same routines as discussed in Section~\ref{sec:teststatcal}, but replacing $\mathbf{S}(\theta_*)$ with $\mathrm{diag}\left(\boldsymbol{\xi} - \bar{\xi}\,\right) \mathbf{S}(\theta_n)$. 

\subsection{Confidence band construction}

As discussed in Section 3, to construct confidence bands, the class $\Theta$ must be assumed to have sufficient structure.
We first discuss the construction of a data-driven approximation to $\Theta$.
Similarly as with our construction of the class of directions $\mathcal{H}$, we define our function class using a basis expansion with an additional smoothness constraint,
\begin{align*}
\Theta_{\zeta} := \left\{ \theta = \textstyle\sum_{j=1}^d a_j\eta_j: \sum_{j=1}^d\frac{a_j^2}{\kappa_j} < \zeta \right\},
\end{align*}
where $\zeta >0$ is a bound on the allowable roughness.
As noted in Section 3, we require that the smoothness parameter $\zeta$ be larger than $J(\theta_0)$ to guarantee that the nominal coverage rate is achieved asymptotically; in practice, we set $\zeta =\zeta_n= J(\theta_n)$.
Coverage can be compromised since $J(\theta_n)$ may be a biased estimator of $J(\theta_0)$ when the tuning parameter $\lambda$ for $\theta_n$ in \eqref{PenOpt} is selected so that $\theta_n$ is optimal with respect to the mean squared error --- \cite{geer2000empirical} provides a comprehensive discussion of key issues in penalized least squares regression.
In simulations, we will see that in the oracle setting,  where $\zeta = J(\theta_0)$, nominal coverage is achieved.
We also examine the effect of data-adaptive selection on coverage. 

We now discuss the computation of confidence bands.
For a norm $\Omega$, let $t_*$ be the $(1-\alpha)$-quantile of the limiting distribution of $\Omega(n^{1/2}\dot{R}_{n,\theta_*})$, which can be approximated via the multiplier bootstrap.
The upper limit of a confidence interval for the evaluation $\theta_0(x_0)$ of $\theta_0$ at a fixed point $x_0$ can be taken to be
\begin{align}
\sup_{a_1,a_2,\ldots,a_d} \left\{ \textstyle\sum_{j=1} a_j \eta_j(x_0) : \sum_{j=1}^d \frac{a_j^2}{\kappa_j} < \zeta,\  \Omega\left(n^{1/2}\dot{R}_{n,\sum_{j=1}^d a_j\eta_j}\right) < t_* \right\},
\label{UpperLimit}
\end{align}
that is, the largest value of $\theta(x_0)$  for $\theta \in \Theta_{\zeta}$ \st{and} such that the test statistic is not sufficiently large to reject  the hypothesis $\theta_0 = \theta$ against its complement $\theta_0\neq \theta$.
The lower confidence limit takes a similar form but replacing supremum with infimum.


The optimization problem can be challenging if the test statistic $\Omega(n^{1/2}\dot{R}_{n, \sum_j a_j \eta_j})$ is not available as a  closed-form function of coefficients $(a_1,a_2,\ldots,a_d)$.
We consider the penalized version of the supremum norm test statistic in \eqref{QuadFormSupNorm} (with $\lambda_1$ and $\lambda_2$ fixed) and the $L_2$ norm test statistic in \eqref{ApproxL2Norm}; both are available as a quadratic form and can be written as $\mathbf{S}\big(\sum_{j=1}^d a_j\eta_j\big)^\top \Pi \mathbf{S}\big(\sum_{j=1}^d a_j\eta_j\big)$, where $\Pi$ is a matrix that does not depend on the coefficients $a_1,a_2,\ldots,a_d$.
For $\mathbf{S}\big(\sum_{j=1}^d a_j\eta_j\big)$ to be available as a closed-form function of $a_1,a_2,\ldots,a_d$, an estimate of the conditional mean of $\theta(X) =  \sum_{j=1}^d a_j\eta_j(X)$ given $W$ must also be available in closed form.
Similarly as in our construction of estimators of $\mu_{h, P_0}$ described in Section 5.1, we use
$\mu_{n, \theta, P_0} := \sum_{j=1}^d a_j \mu_{n, \eta_j, P_0}$.
With this construction, $\mathbf{S}$ is linear in $a_1,a_2,\ldots,a_d$, and so, the optimization problem \eqref{UpperLimit} is a quadratically constrained quadratic program and can be solved using interior point methods.
Many software packages include implementations for this type of problem --- see, e.g., CVXR  for an implementation in R \citep{fu2017cvxr}.

The last main challenge concerns selection of the class $\mathcal{H}$ of directions.
To construct confidence bands of optimal width, we must have optimal power to reject each false hypothesis $H: \theta_0 = \theta$.
The class of directions that provides the optimal test, however, depends on the null hypothesis --- when the distance between $\theta$ and $\theta_0$ is larger, the class of directions should also be larger.
While the norm $\Omega$ in \eqref{UpperLimit} should therefore depend on $\theta$, allowing this dependence can create computational difficulties.
We instead take the simple approach of using the same class $\mathcal{H}_\gamma$ to perform all tests.
We set $\gamma =\gamma_n= J(\theta_n)V_n^2(\theta_n)$ so that the restricted score test is well-powered against relatively flat nulls when $\theta_0$ is non-smooth.

\section{Results from simulation studies}\label{sec:sims}

In this section, we examine the behavior of our proposed methodology in a simulation study.
The objectives of this simulation study are to demonstrate that our proposed test provides nominal type I error control and coverage, and to examine how the selection of the norm $\Omega$ influences statistical power and the width of resulting confidence bands.

\subsection{Example 1: nonparametric mean regression}

We first consider the nonparametric regression setting.
We generate synthetic data from the model $Y = \theta_0(X) + \epsilon$,
where we set the regression function $\theta_0$ to be
\begin{align}
\theta_0(x) = \sin[\pi x^2\text{sign}(x)]
\label{theta0sim}
\end{align}
and we draw independently $X$ from a uniform distribution on $(-1,1)$ and $\epsilon$ from a $N(0,9)$ distribution.
Under these settings, we generate 800 synthetic data sets for $n \in \{100, 500, 1000, 2000\}$.

We compare the restricted score test using the $L_2$ norm and the quadratic form approximation of the supremum norm described in Section 5.2.
For confidence band construction, we also consider both known and estimated smoothness of $\theta_0$.
We construct $\mathcal{H}$ from a Sobolev basis with $d = 50$ basis functions.
For each application of the multiplier bootstrap, we generate 1000 bootstrap samples.

To assess the statistical power, we test the null hypothesis $H_0:\theta_0\equiv 0$, and to assess type I error rate control, we test the null at the true value of $\theta_0$.
In both cases, we use significance level $\alpha = 0.05$.
To assess coverage, we select 50 evenly spaced points on the interval $[-1, 1]$ and determine if the evaluation of the true regression function at each point lies within the confidence band.
We summarize the width of the band by calculating the average width at these 50 points.

We compare our restricted score test with the debiased local polynomial regression estimator of \cite{calonico2018effect}, implemented in the publicly available R package \texttt{nprobust} \citep{calonico2019nprobust}.
The \texttt{nprobust} package is designed for pointwise inference, so we repurposed the method for hypothesis testing and uniform interval construction as we now describe. For a fixed sequence of points $x_1,x_2,\ldots,x_k$, the \texttt{nprobust} package outputs an estimate $\boldsymbol{\check{\theta}} := (\check{\theta}(x_1),\check{\theta}(x_2) \ldots, \check{\theta}(x_k))$ of $\boldsymbol{\theta}_0 := (\theta_0(x_1),\theta_0(x_2), \ldots, \theta_0(x_k))$ that approximately satisfies  $\boldsymbol{\check{\theta}} \sim
N(
\boldsymbol{\theta}_0,
C_1C_2C_1)$, where  $C_1:=\text{diag}\{\text{sd}( \check{\theta}(x_1) ),\text{sd}( \check{\theta}(x_2) ),\ldots,\text{sd}( \check{\theta}(x_k))\}$ and $C_2 := \text{corr}(\check{\theta}(x_1),\check{\theta}(x_3),\ldots,\check{\theta}(x_k))$, for sufficiently large $n$.
We use a confidence band of the form
\begin{align*}
\left(\check{\theta}(x_j) - m_{1-\alpha}\, \text{sd}( \check{\theta}(x_j)),\ \check{\theta}(x_j) + m_{1-\alpha}\,\text{sd}( \check{\theta}(x_j) ) \right),\ \ j = 1,2,\ldots,k,
\end{align*}
where $m_{1-\alpha}$ is the $(1-\alpha)$-quantile of the distribution for the maximum absolute value of a Gaussian random vector with mean zero and variance $C_2$.
We test $H_0:\theta_0 = \theta_*$ by verifying whether $\theta_*$ resides within the interior of the confidence band, and reject the null hypothesis whenever
\begin{equation*}
\max_{j \in \{1,2,\ldots,k\}} \frac{|\check{\theta}(x_j) - \theta_*(x_j)|}{\text{sd}(\check{\theta}(x_j))} > m_{1-\alpha}\ .
\end{equation*}

\begin{figure}[!ht]
\centering
\includegraphics[scale=.825]{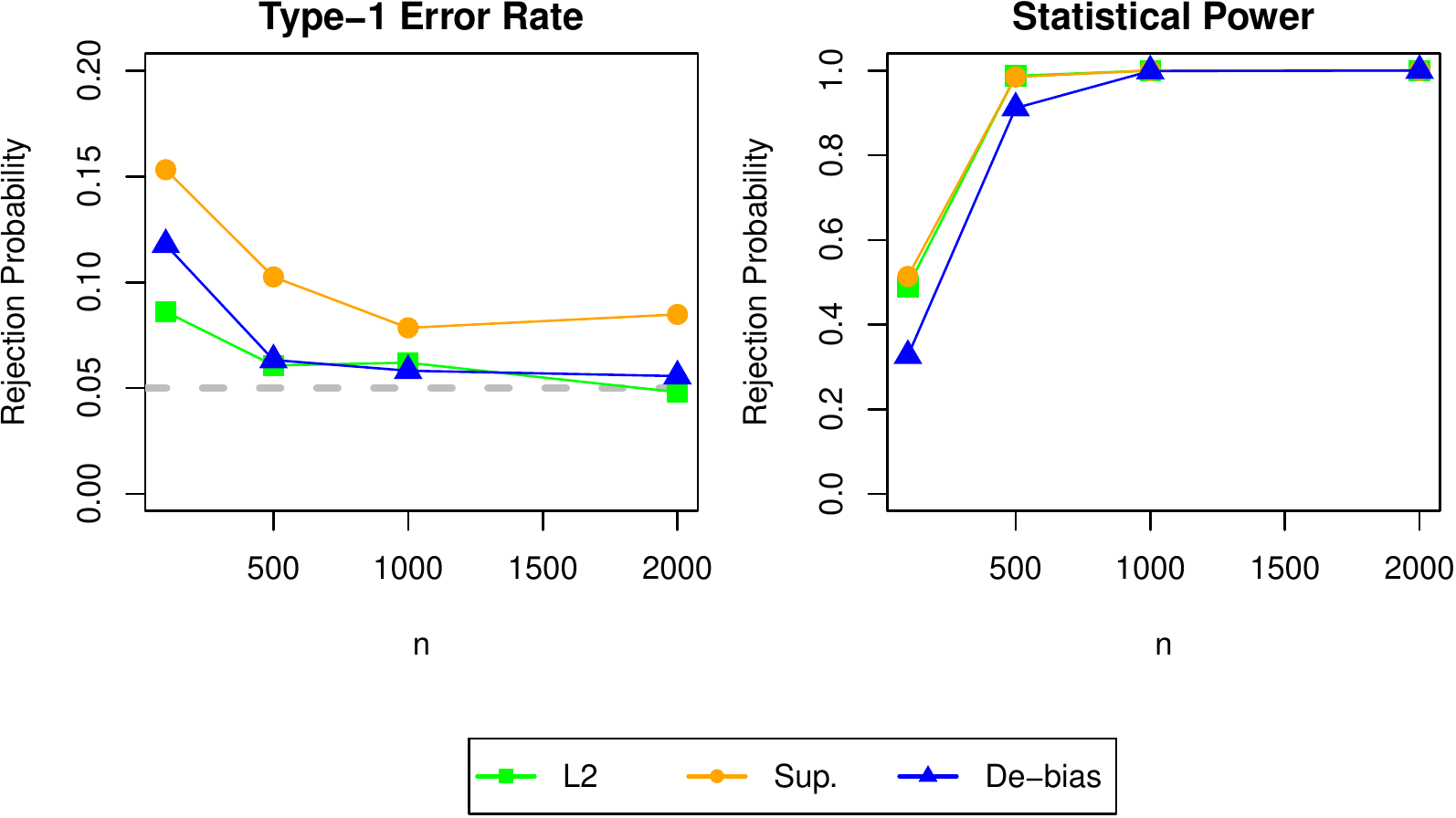}
\caption{Monte Carlo estimates of type-1 error rate (left) and statistical power (right) in the regression setting.  The dashed gray line indicates the significance level $\alpha = .05$.}
\label{fig:RegType1Power}
\end{figure}

\begin{figure}[!hb]
\centering
\includegraphics[scale=.825]{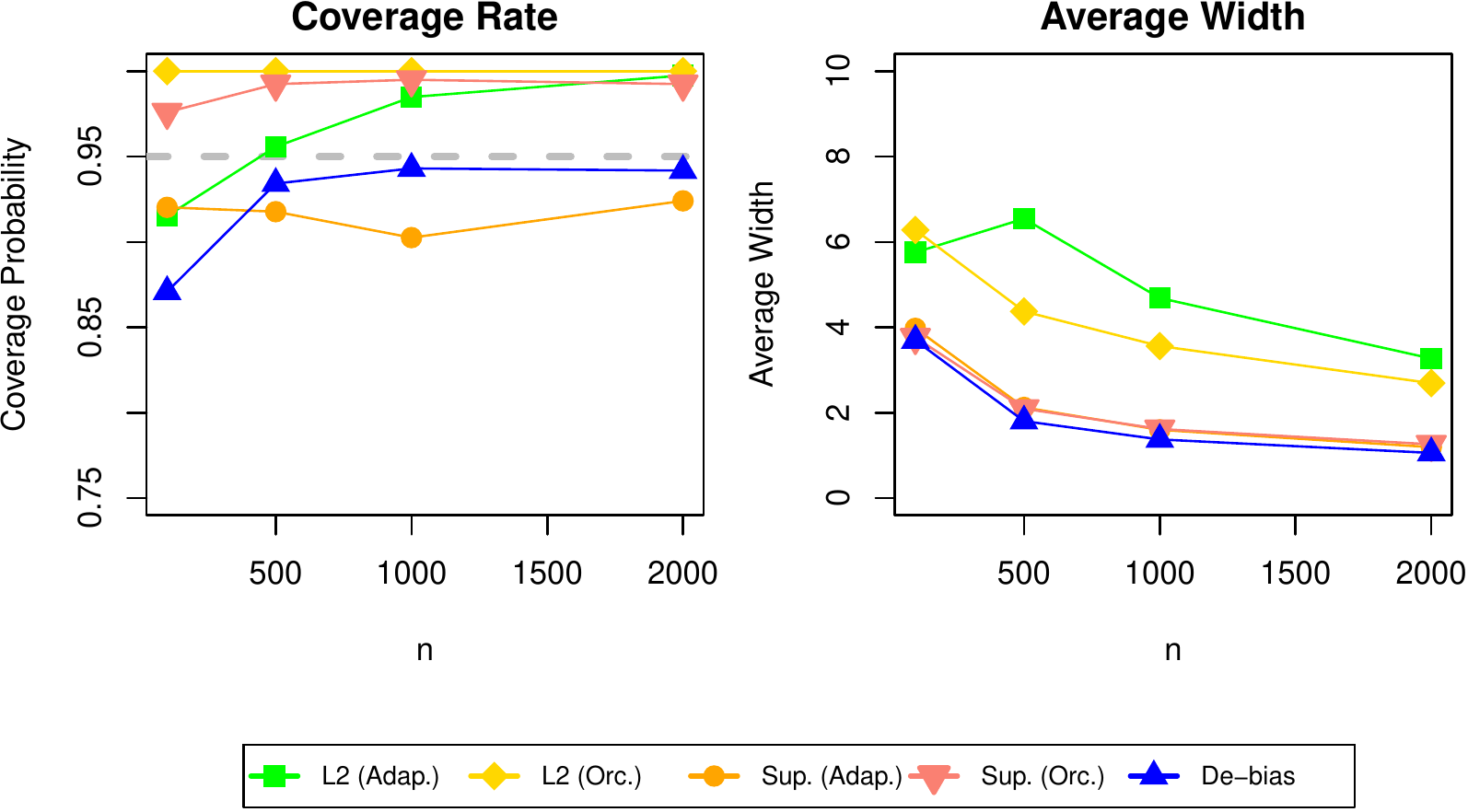}
\caption{Monte Carlo estimates of the coverage probability (left) and average width of confidence band (right) in the regression setting.  The dashed gray line indicates the nominal coverage rate $.95$.}
\label{fig:RegCoverWidth}
\end{figure}

\begin{figure}[!ht]
\centering
\includegraphics[scale=.825]{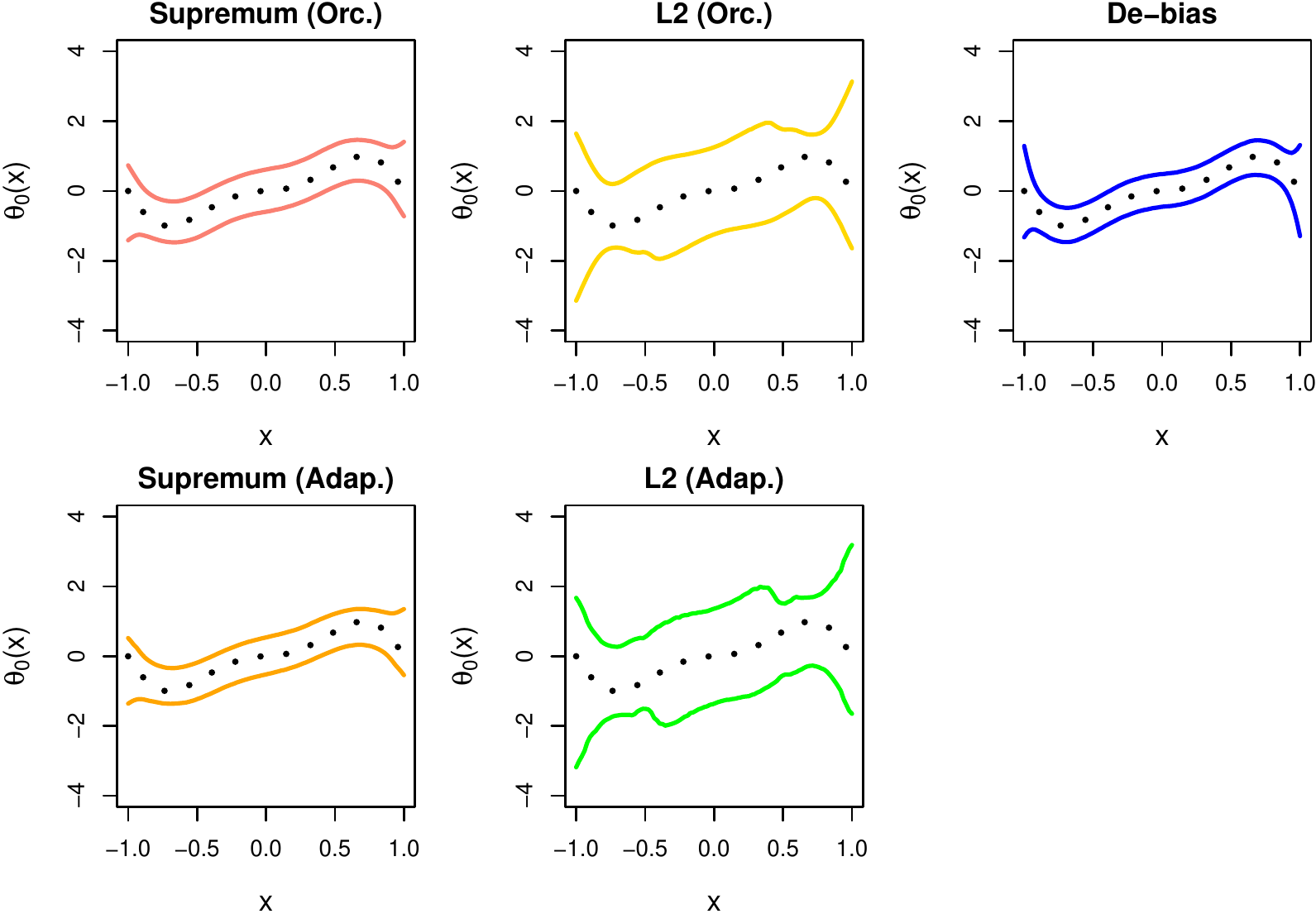}
\caption{Median upper and lower limits of confidence bands with $n = 2000$ in the regression setting.  The dotted black line represents the true risk minimizer $\theta_0$.}
\label{fig:RegShape}
\end{figure}

We summarize results in Figures \ref{fig:RegType1Power}, \ref{fig:RegCoverWidth} and \ref{fig:RegShape}.
The restricted score test provides nominal type I error control in moderate sample sizes when using  the $L_2$ norm and is slightly anti-conservative when using the penalized supremum norm.  
The restricted score test is also well-powered against the null hypothesis $H_0: \theta_0 \equiv 0$; the power is comparable for the $L_2$ and penalized supremum norms, and both choices offer a modest improvement over the debiasing approach.
The confidence bands constructed via the restricted score test exceed the nominal coverage rate when the oracle choice smoothness $\zeta = J(\theta_0)$ is supplied.
This is unsurprising, as our construction in Section 5.2 only guarantees that the coverage rate will be \textit{no smaller than} $(1-\alpha)$.
A confidence set for the collection of all functionals of $\theta_0$ should achieve the nominal coverage rate, but we only consider the  subset of evaluation functionals.
When using the smoothness of the estimate $\zeta =\zeta_n= J(\theta_n)$, the coverage rate remains close to the nominal level when the supremum norm is used, and exceeds the nominal rate when the $L_2$ norm is used.
The debiasing approach provides the narrowest confidence bands, followed closely by the penalized supremum norm with adaptively-selected smoothness $\zeta_n$ and the penalized supremum norm with oracle smoothness.
The median upper and lower confidence limits for all methods with $n = 2000$ are provided in Figure \ref{fig:RegShape}.
We find that the confidence bands are able to best capture the shape of the true risk minimizer $\theta_0$ when the supremum norm is used.

\subsection{Example 2: partially additive mean regression}

Our simulation design for the partially additive mean model is similar to the design used in the nonparametric regression setting.
We let $W := (W_1, W_2)$ be a vector of two independent uniform random variables on $(-1,1)$ and subsequently generate $X=\frac{1}{3}W_1 + \frac{1}{3}\sin(\pi W_2) + \Delta$,
where $\Delta$ follows a uniform distribution on $(-\frac{1}{3}, \frac{1}{3})$.
By construction, $X$ has support $(-1, 1)$.
We then generate $Y= f_0(W) + \theta_0(X) + \epsilon$, 
where the nuisance function $f_0$ is defined pointwise as 
\begin{align*}
f_0(w_1, w_2) =- 4 \left[\frac{\exp(5 w_1)}{1 + \exp(5 w_1)} - \frac{1}{2}\right] - 2\, \text{sign}(w_2) w_2^2\ ,
\end{align*}
the parameter $\theta_0$ of interest  is defined as in \eqref{theta0sim}, and $\epsilon$ is a  $N(0,9)$ random variable independent of $(W,X)$.

We compare our restricted score test under the same settings as in the nonparametric regression example, with the exception that we construct $\mathcal{H}$ using a smaller set of $d = 10$ basis functions.
We use the highly adaptive lasso \citep{benkeser2016highly} to estimate the conditional mean functions $\mu_{Y, P_0}$, $\mu_{\theta_*, P_0}$ and $\mu_{h, P_0}$.
To the best of our knowledge, construction of uniform confidence bands in partially additive models has not been studied, so we do not compare with a competing method.

\begin{figure}[!hb]
\centering
\includegraphics[scale=.825]{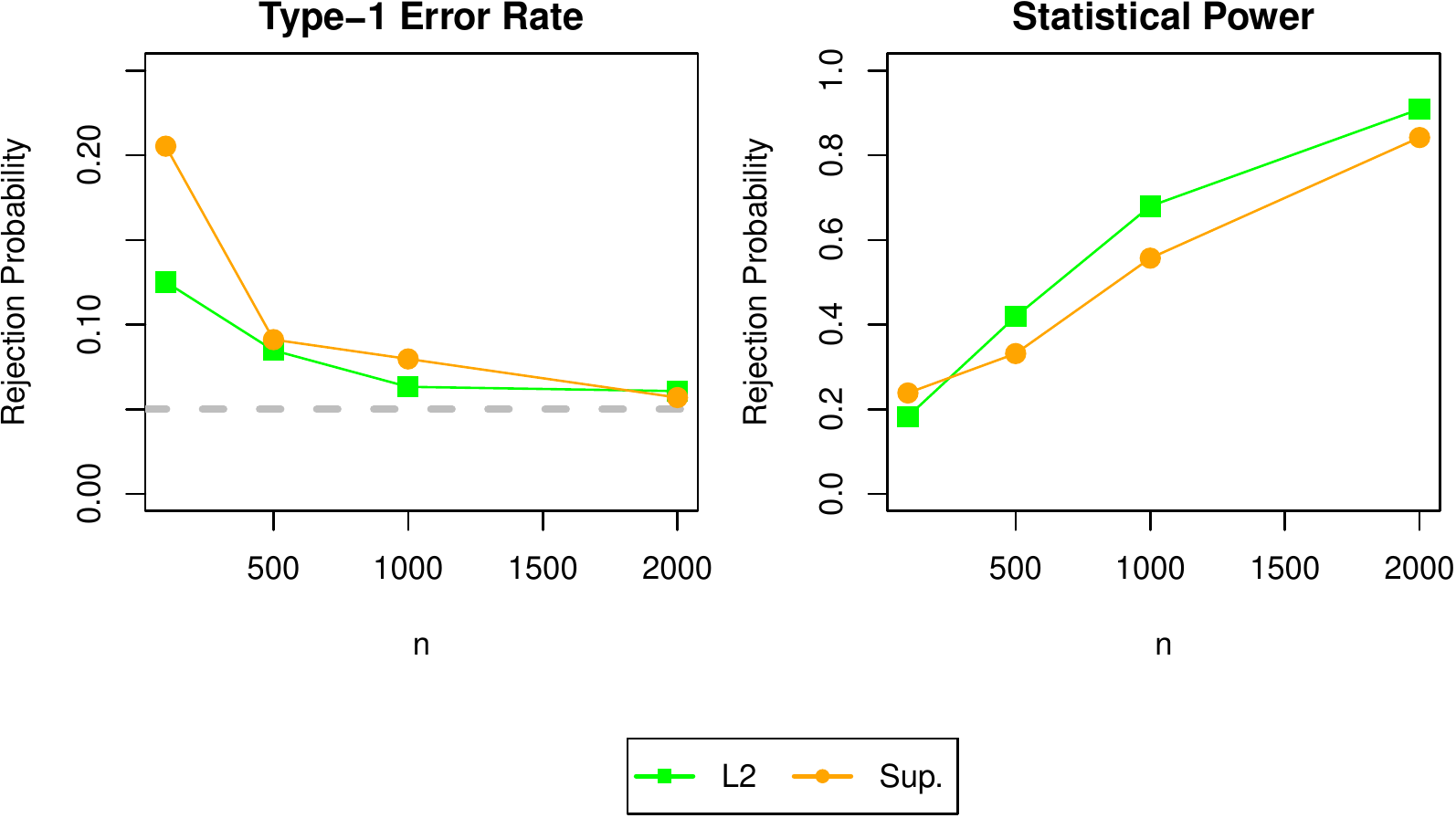}
\caption{Monte Carlo estimates of type-1 error rate (left) and statistical power (right) for the partially additive model.  The dashed gray line indicates the significance level $\alpha = .05$.}
\label{fig:PamType1Power}
\end{figure}

\begin{figure}[!ht]
\centering
\includegraphics[scale = .825]{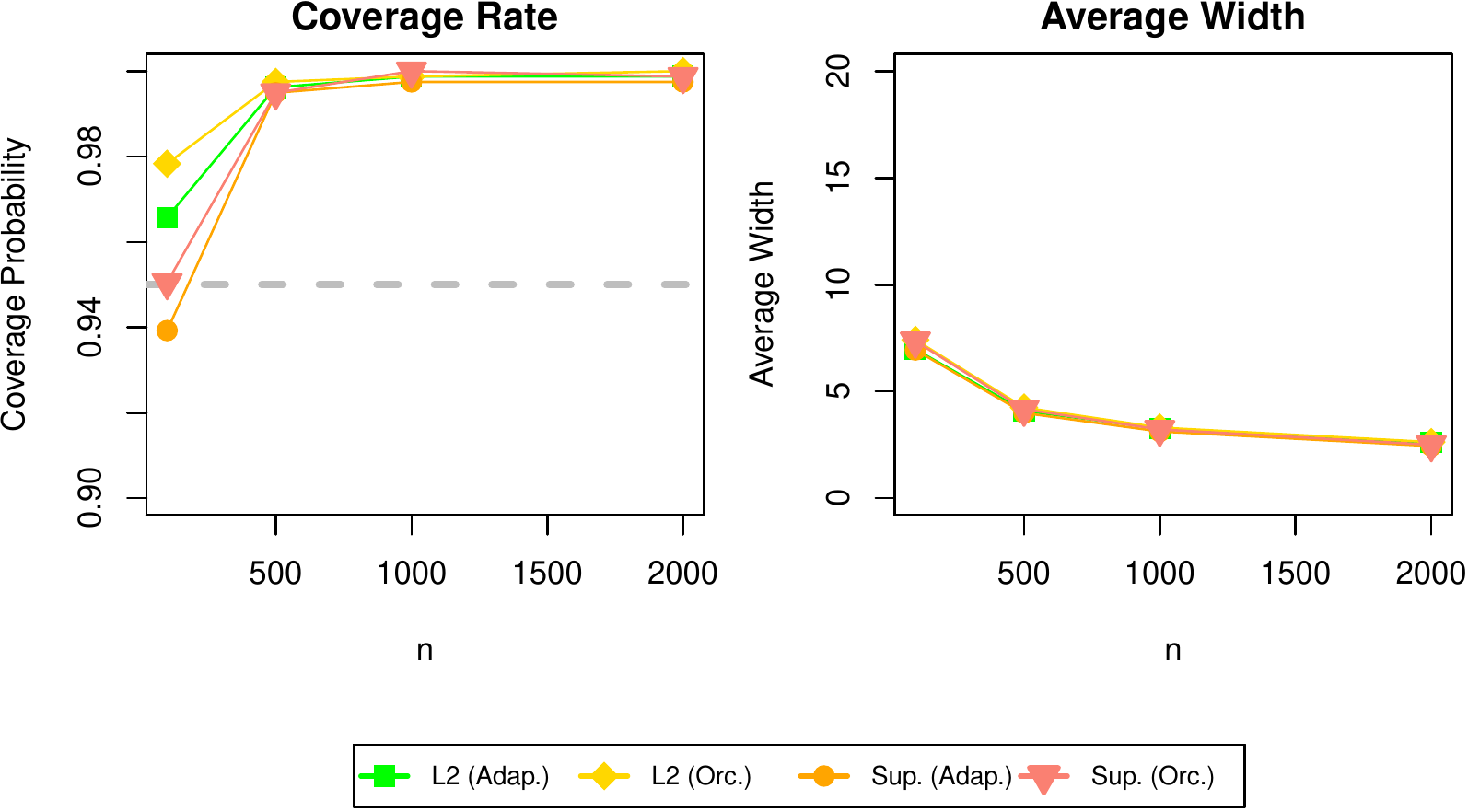}
\caption{Monte Carlo estimates of the coverage probability (left) and average width of confidence band (right) for the partially additive model.  The dashed gray line indicates the nominal coverage rate $.95$.}
\label{fig:PamCoverWidth}
\end{figure}

\begin{figure}[!hb]
\centering
\includegraphics[scale = .825]{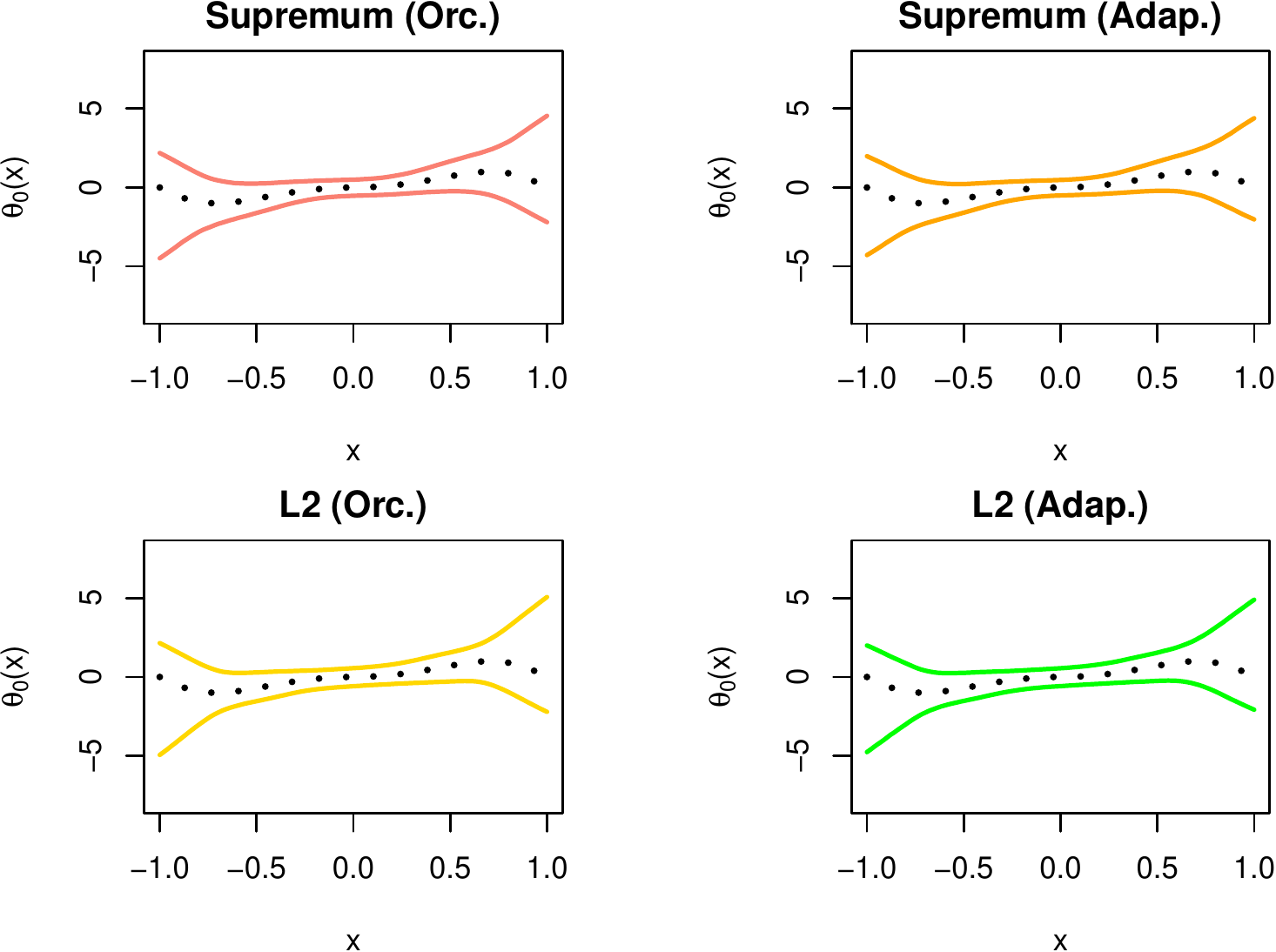}
\caption{Median upper and lower limits of confidence bands with $n = 2000$ for the partially additive model.}
\label{fig:PamShape}
\end{figure}

Simulation results are presented in Figures \ref{fig:PamType1Power}, \ref{fig:PamCoverWidth} and \ref{fig:PamShape}.
Similarly as in the nonparametric regression setting, we find that the restricted score test achieves nominal control of the type I error rate in large sample sizes, and the statistical power is comparable for the both $L_2$ and penalized supremum norms.
For all methods considered, the confidence bands exceed the nominal coverage rate, suggesting that estimating the smoothness of $\theta_0$ does not seriously compromise coverage guarantees.
The average width of the confidence bands obtained using the penalized supremum norm and the $L_2$ norm are similar, and the bands are generally able to capture the shape of $\theta_0$.
The width is larger in the tails because $X$ is not uniformly distributed --- its distribution is more tightly concentrated around zero.

\section{Results from the 1987 National Medical Expenditure Survey}\label{sec:realdata}

We apply our method to data from the 1987 National Medical Expenditure Survey, extracted by \cite{johnson2003disease}.
These data include information about smoking behaviors and medical expenditures in a sample of 9,708 US citizens.
The objective of this analysis is to assess the association between tobacco exposure, measured in pack-years (the number of cigarette packs an individual smoked per day times the number of years they smoked), and medical expenditure.

We fit a partially additive mean model in which where the outcome $Y$ is the total medical expenditure, the exposure $X$ is the natural log of the number of pack-years, and the adjustment covariate vector $W$ includes: participant age at the time of the survey, age at initiation of smoking, gender, marital status, education level, census region, and socioeconomic status.
We are interested in testing the null hypothesis that there is no association between pack-years smoked and medical expenditures ($\theta_0 \equiv 0$) and in constructing a confidence band for $\theta_0$.
We  apply the restricted score test using the approximate supremum norm for testing and confidence band construction.
As in the simulation study, we construct the class of directions $\mathcal{H}$ using a Sobolev basis with $d = 10$ basis functions.

\begin{figure}[!h]
\centering
\includegraphics[scale = .75]{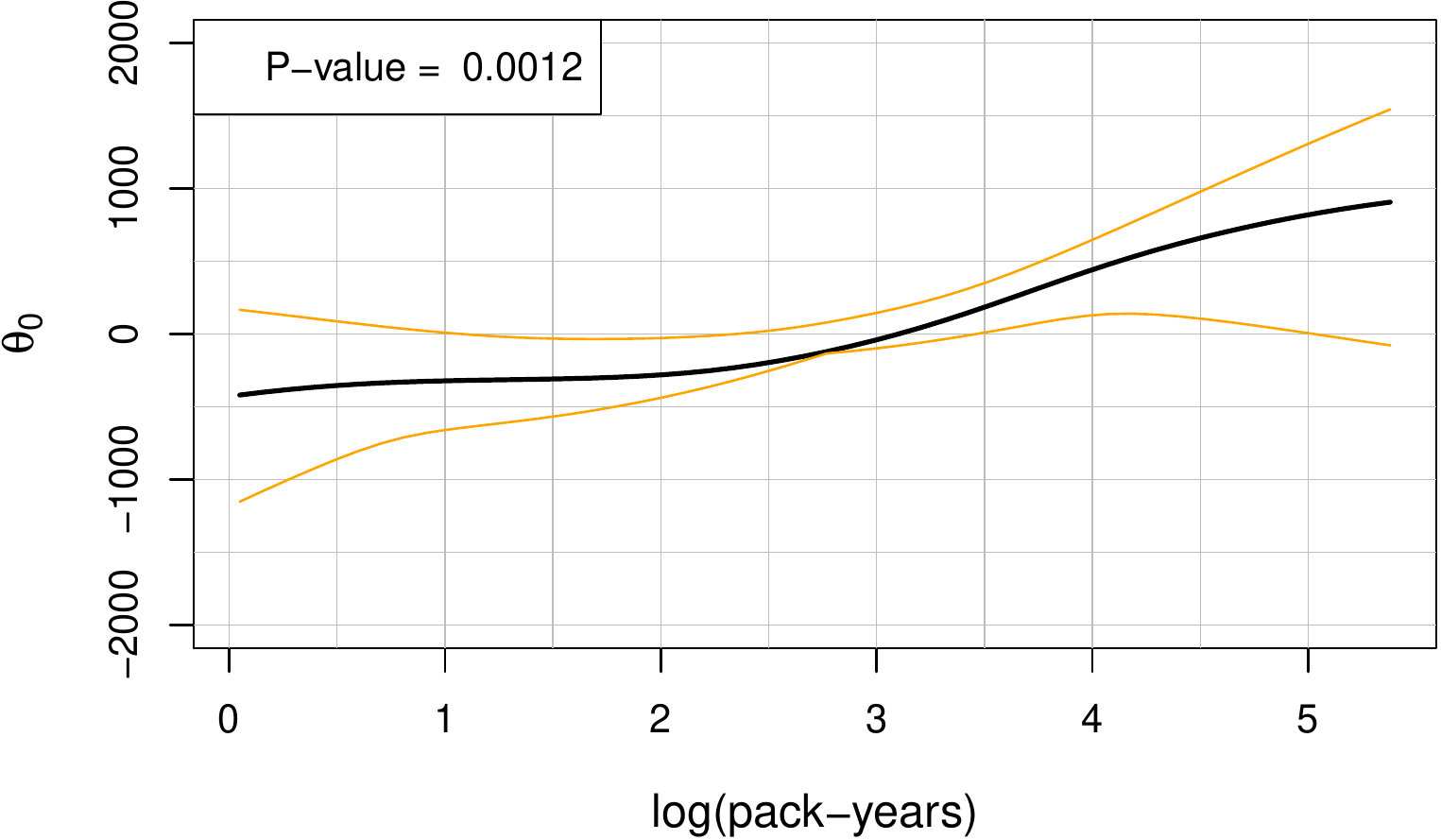}
\caption{Estimate and confidence band for partially additive model fit to the 1987 National Medical Expenditure Survey data.}
\label{fig:NMESResults}
\end{figure}

Results from the analysis are summarized in Figure \ref{fig:NMESResults}.
We find strong evidence of an association between pack-years and medical expenditure, with p-value  $p=0.001$ based on 10,000 bootstrap samples.
The resulting estimate of the regression function suggests that a small amount of smoking (less than 2 $\log$-pack-years) does not substantially increase average medical expenditure.
The effect of smoking is more apparent for those who have had a moderate or large amount of exposure to smoking.

\section{Discussion}\label{sec:disc}

We have introduced a general approach for hypothesis testing and constructing confidence bands for infinite-dimensional parameters in nonparametric and semiparametric statistical models.
Our framework is applicable to any function-valued parameter that can be expressed as a risk minimizer.
While we consider the nonparametric and semiparametric partially additive mean regression models as examples in this paper, the framework can be useful in other widely relevant applications.
For example, in causal inference, both the conditional average treatment effect curve and the causal dose-response function are parameters tat can be expressed expressed as risk minimizers, so that inference can be conducted using the restricted score test.

There remain several important issues that must be clarified through additional research.
In order to guarantee adequate coverage for resulting confidence bands, our approach requires an upper bound for the smoothness of the true risk minimizer, and this requirement can be seen as a limitation of the method.
However, we found in simulations that using a naive plug-in estimator may work well in practice, even if such estimators are not necessarily consistent.
Regardless, our confidence band with estimated smoothness retains a meaningful interpretation as an envelope containing a set of smooth functions consistent with the observed data.
Additional work is also required to provide further guidance on the selection of the class of directions $\mathcal{H}$ and the choice of norm $\Omega$.
Though these choices do not affect type I error control, they can influence statistical power and the width of resulting confidence bands.
For the purposes of this paper, we have only provided some heuristic guidance for selecting a class of directions, and leave more extensive and rigorous studies of this question to future work.

\section{Acknowledgements}

The authors gratefully acknowledge the support of the NSF Graduate Research Fellowship Program under grant DGE-1762114 as well as NSF grant DMS-1561814, NHLBI grant HL137808, and NIH grant R01-GM114029.
Any opinions, findings, and conclusions or recommendations expressed in this material are those of the authors and do not necessarily reflect the views of the funding agencies.



%

\newpage
\singlespacing
\bibliography{restgradtest-bib}

\newpage
\doublespacing

\newpage

\newpage

\newpage

\newpage

\section*{Supplementary material}\vspace{.1in}


\subsection*{PART I: Additional technical details}\vspace{.1in}

\noindent \textbf{Estimation of G\^{a}teaux derivative in Example 2}\vspace{.1in}

\noindent Recall the form of the G\^ateaux derivative of the risk functional for the partially additive mean regression model in \eqref{PAM-GD}.
The next result provides conditions under which the plug-in estimator in  \eqref{PAM-Plug} is uniformly asymptotically linear with the efficient influence function defined in \eqref{PAM-EIF}. Below, we denote by $\phi_{n,\theta_*}(\cdot; h)$ the estimated influence function \begin{align*}
z=(w,x,y)\mapsto \phi_{n,\theta_*}(z; h):=\{y - \mu_{n, Y, P_0}(w) + \mu_{n, \theta_*, P_0}(w)- \theta_*(x)\}\{h(x) - \mu_{n, h, P_0}(w) \} - \dot{R}_{n, \theta_*}(h)\ .
\end{align*}

\begin{result}
Suppose that there exists a $P_0$-Donsker class $\Phi$ such that $\phi_{P_0, \theta_*}(\cdot; h)$ and $\phi_{n, \theta_*}(\cdot;h)$ are both in $\Phi$  for all $h \in \mathcal{H}$ with probability tending to one.
Furthermore, suppose that the rate conditions $\int \left\{ \mu_{n,Y,P_0}(w) - \mu_{Y,P_0}(w) \right\}^2 dP_0(w) = o_P(n^{-1/2})$, $\int \left\{ \mu_{n,\theta_*,P_0}(w) - \mu_{\theta_*,P_0}(w) \right\}^2 dP_0(w) = o_P(n^{-1/2})$, and
    \[\sup_{h \in \mathcal{H}}\int \left\{\mu_{n,h,P_0}(w) - \mu_{h,P_0}(w) \right\}^2 dP_0(w) = o_P(n^{-1/2})\]
are satisfied.
Then, it follows that $\sup_{h \in \mathcal{H}} |r_n(h)| = o_P(n^{-1/2})$, where $r_n(h)$ is the remainder term in \eqref{AsympLin}.
\end{result}

\noindent \textbf{Proof of Result 1.}\\
\noindent The remainder term can be written as $r_n(h) = A_n(h) + B_n(h)$, where we define
\begin{align*}
   A_n(h) &:=
   \int \left\{\phi_{n,\theta_*}(z; h) - \phi_{P_0,\theta_*}(z; h)\right\}d(P_n - P_0)(z)
    \\
    B_n(h) &:= \int \left\{\phi_{n, \theta_*}(z;h) - \phi_{P_0, \theta_*}(z;h)\right\}dP_0(z) + \dot{R}_{n, \theta_*}(h) - \dot{R}_{0, \theta_*}(h).
\end{align*}
For the first term, it is shown in the proof of Lemma 19.26 of \cite{van2000asymptotic} that $\sup_{h \in \mathcal{H}} |A_n(h)| = o_P(n^{-1/2})$ in view of the Donsker class condition and uniform consistency of the nuisance parameter estimator.
The second term can be expressed as
\begin{align*}
B_n(h) = \int\{\mu_{Y, P_0}(w) - \mu_{n,Y,P_0}(w) + \mu_{n,\theta_*,P_0}(w)- \mu_{\theta_*, P_0}(w)\}
\left\{\mu_{n,h,P_0}(w) - \mu_{h, P_0}(w)\right\} dP_0(z).
\end{align*}
By an application of the Cauchy-Schwarz inequality, and in view of the rate conditions, it follows that $\sup_{h \in \mathcal{H}}|B_{n}(h)| = o_P(n^{-1/2})$.
This completes the proof in view of the triangle inequality.

\subsection*{PART II: Proof of lemma and theorems}\vspace{.1in}

\noindent \textbf{Proof of Lemma 1.}\\
The result follows immediately from Slutsky's theorem  (see, e.g., Theorem 7.15 of \citealp{kosorok2008introduction}).

\noindent \textbf{Proof of Theorem 1.}\\
Below, denote by $\mathcal{B}$ the collection containing each bounded Lipschitz functionals $g: \ell^\infty(\mathcal{H}) \to [-1, 1]$ with Lipschitz constant 1, that is, satisfying that $|g(a_1) - g(a_2)| \leq \|a_1 - a_2\|_{\mathcal{H}}$ for all $a_1, a_2 \in 
\ell^\infty(\mathcal{H})$. Define the real-valued random functionals $\Gn:h\mapsto n^{-1/2}\sum_{i=1}^{n}\xi_i \phi_{f_n,\theta_*}(Z_i;h)$ and $\Gzero:h\mapsto n^{-1/2}\sum_{i=1}^{n}\xi_i \phi_{f_0,\theta_*}(Z_i;h)$ defined on $\mathcal{H}$. We will show that 
\begin{align}\label{eqn:thm1claim}
    \sup_{g \in \mathcal{B}}  \left|E_{\xi}\left[g(\Gn)\right] - 
    E_0\left[g(\G)\right] \right|
\end{align}
converges to zero in outer probability, where $E_\xi$ denotes expectation over the distribution of $\xi_1,\xi_2,\ldots,\xi_n$, which implies the desired result since convergence of the expectation of bounded Lipschitz functions of a stochastic process is equivalent to weak convergence in view of the Portmanteau lemma \citep[see, e.g., Lemma 18.9 of][]{van2000asymptotic}. 

First, by the triangle inequality, we note that
\begin{align*}
    \sup_{g \in \mathcal{B}}  \left|E_{\xi}\left[g(\Gn)\right] - 
    E_0\left[g(\G)\right] \right|\ \leq\ 
    \sup_{g \in \mathcal{B}} \left|E_{\xi}\left[g(\Gzero)\right] - 
    E_0 \left[g(\G)\right] \right| + \sup_{g \in \mathcal{B}} \left|E_{\xi}\left[g(\Gn)-
    g(\Gzero)\right] \right|.
\end{align*}
We define the function $\bar{g}:\ell^{\infty}(\mathcal{H}) \rightarrow \mathbb{R}$ pointwise as $\bar{g}(u):= \min\left(1,\|u\|_{\mathcal{H}}\right)$. 
Since $|g(u_1) - g(u_2)| \leq \min(2, \|u_1 - u_2\|_\mathcal{H})$ for any $g \in \mathcal{B}$ and $u_1, u_2 \in \ell^\infty(\mathcal{H})$, we get that
\begin{align*}
    \sup_{g \in \mathcal{B}} \left|E_{\xi}\left[g(\Gn)-
    g(\Gzero)\right] \right|\ \leq\ 2\left|E_{\xi}\left[\bar{g}(\Gn-\Gzero)\right] \right|.
\end{align*}    
Defining for each $z\in\mathcal{Z}$ the random functionals $\varphi_{n}(z):h\mapsto \phi_{f_n,\theta_*}(z;h)-\phi_{f_0,\theta_*}(z;h)$ and $\overline{\varphi}_n(z):=\varphi_n(z)-\int \varphi_n(z)dP_0(z)$, we use the triangle inequality again to establish that
\begin{align*}    
    \sup_{g \in \mathcal{B}}  \left|E_{\xi}\left[g(\Gn)\right] - 
    E_0\left[g(\G)\right] \right|\ &\leq\ \sup_{g \in \mathcal{B}}  \left|E_{\xi}\left[g(\Gzero)\right] - 
    E_0\left[g(\mathbb{G})\right] \right|+2(A_{n}+B_n)\ ,
\end{align*}where we have defined
\begin{align*}
    A_n\ &:=\ \left| E_{\xi}\left[\bar{g}\left(n^{-1/2} \textstyle\sum_{i=1}^n \xi_i\varphi_n(Z_i)\right)-\bar{g}\left(n^{-1/2}\textstyle\sum_{i=1}^n\xi_i\overline{\varphi}_n(Z_i)\right)\right] \right|
    \\
    B_n\ &:=\ \left| E_\xi\left[\bar{g}\left(n^{-1/2}\textstyle\sum_{i=1}^n\xi_i\overline{\varphi}_n(Z_i)\right)\right] \right|\ .
\end{align*}The first summand above converges to zero in outer probability by Theorem~2.9.6 of \cite{van1996weak}.
We find that $A_n\geq 0$ tends to zero in probability by observing that 
\begin{align*}
   A_n\ &\leq\ \sup_{g \in \mathcal{B}}\left| E_{\xi}\left[g\left(n^{-1/2} \textstyle\sum_{i=1}^n \xi_i\varphi_n(Z_i)\right)-g\left(n^{-1/2}\textstyle\sum_{i=1}^n\xi_i\overline{\varphi}_n(Z_i)\right)\right] \right|\\
   &\leq\ \sup_{h \in \mathcal{H}}\left|\int\varphi_n(z)(h)dP_0(z)\right| E_{\xi} \left|n^{-1/2}\textstyle\sum_{i=1}^n \xi_i\right|\\
   &\leq\ \left[\sup_{h \in \mathcal{H}}\int\varphi_n(z)(h)^2dP_0(z)\,E_{\xi} \left(n^{-1}\textstyle\sum_{i=1}^n \xi^2_i\right)\right]^{1/2}=\ \left[\sup_{h \in \mathcal{H}}\int\varphi_n(z)(h)^2dP_0(z)\right]^{1/2}=\ o_P(1)\ ,
\end{align*}where the first inequality holds because $\bar{g}$ resides in $\mathcal{B}$, the second inequality holds in view of the Lipschitz property, the third inequality follows from the Cauchy-Schwarz inequality and the fact that $\xi_1,\xi_2,\ldots,\xi_n$ are independent and have mean zero, the first equality holds because $\xi_1,\xi_2,\ldots,\xi_n$ have unit second moment, and the last statement follows by assumption. Finally, we focus on the term $B_n\geq 0$. 
By the triangle inequality and the fact that $\bar{g}\in\mathcal{B}$, we have that $B_n\leq B_{1n}+B_{2n}$, where
\begin{align*}
B_{1n}\ &:=\ \sup_{g\in\mathcal{B}}\left|E_\xi\left[g\left(n^{-1/2}\textstyle\sum_{i=1}^{n}\xi_i\overline{\varphi}_n(Z_i)\right)\right]-E_0\left[g\left(n^{1/2}\textstyle\int \varphi_n(z)d(P_n-P_0)(z)\right)\right]\right|\\
B_{2n}\ &:=\ \left|E_0\left[\bar{g}\left(n^{1/2}\textstyle\int \varphi_n(z)d(P_n-P_0)(z)\right)\right]\right| \ .
\end{align*}We define $\nu(z): (h,f) \mapsto \phi_{f,\theta_*}(z;h) - \phi_{f_0, \theta_*}(z;h)$ and $\bar{\nu}(z): (h,f) \mapsto \nu(z) - \int \nu(z) dP_0(z)$ as real-valued functionals over $\mathcal{H} \times \mathcal{F}$.
Define $\mathcal{F}_{\delta} := \{f \in \mathcal{F}: \|f - f_0\|_{\mathcal{F}} < \delta\}$, and
let $\mathcal{D}$ be the collection containing each bounded Lipschitz functional $q: \ell^\infty\left(\mathcal{H}\times \mathcal{F}_{\delta}\right) \to [-1, 1]$ with Lipschitz constant 1.
For any $\delta>0$, since $\phi_{f_n,\theta_*}(\cdot; h) - \phi_{f_0, \theta_*}(\cdot; h)$ is in $\Phi_\delta$ for each $h \in \mathcal{H}$ with probability tending to one, we have that
\begin{align*}
    B_{1n}\ \leq\ \sup_{q\in\mathcal{D}}\left|E_\xi\left[q\left(n^{-1/2}\textstyle\sum_{i=1}^{n}\xi_i\overline{\nu}(Z_i)\right)\right]-E_0\left[q\left(n^{1/2}\textstyle\int \nu(z)d(P_n-P_0)(z)\right)\right]\right|
\end{align*}with probability tending to one.
Since $\Phi_{\delta}$ is $P_0$-Donsker for small enough $\delta>0$, we find that this upper bound for $B_{1n}$ tends to zero in outer probability in view of Theorem 2.9.6 of \cite{van1996weak}.
Finally, we argue that $B_{2n}$ tends to zero (deterministically). Defining the random variable $G_n:=\bar{g}\left(n^{1/2}\int \varphi_n(z)d(P_n-P_0)(z)\right)$, we note that the sequence $\{G_1,G_2,\ldots\}$  is uniformly bounded by one. Furthermore, because $0\leq G_n\leq \sup_{h\in\mathcal{H}}|n^{1/2}\textstyle\int \varphi_n(z)(h)d(P_n-P_0)(z)|$, $G_n$ tends to zero in probability provided \begin{align*}
    &\sup_{h \in \mathcal{H}}\left|n^{1/2}\int\varphi_n(z)(h)d(P_n - P_0)(z) \right| = o_P(1)\ ,
\end{align*}in which case it follows that $B_{2n}=|E_0(G_n)|$ tends to zero as well. This condition is shown in the proof of Lemma 19.26 of \cite{van2000asymptotic}, thereby completing the proof.

\noindent \textbf{Proof of Theorem 2.}

\noindent For conciseness, for any given probability $P$ and $P$-integrable function $f$, we denote $\int f(z)dP(z)$ by the shorthand notation $Pf$ below, and we write $\mathbb{G}_n:=n^{1/2}(P_n-P_0)$. We begin by proving (a). By application of the continuous mapping theorem and Slutsky's theorem, the result follows if we can show that
\begin{equation}
\sup_{h \in \mathcal{H}_n} \mathbb{G}_n\phi_h = \sup_{h \in \mathcal{H}} \mathbb{G}_n\phi_h+o_P(1)\ .
\label{Lemma1eq1}
\end{equation}
By \eqref{hausdorff}, there exists a deterministic sequence $\epsilon_n\downarrow 0$ such that
\begin{align*}
P_0\left(\sup_{h_1 \in \mathcal{H}} \inf_{h_2 \in \mathcal{H}_{n}}P_0(\phi_{h_1}-\phi_{h_2})^2 > \epsilon_n \right) \longrightarrow 0\ .
\end{align*}
By the definition of the supremum and infimum, for each $n$, there exist random functions $h_{1,n}\in\mathcal{H}$ and $h_{2,n}\in\mathcal{H}_n$ such that $
\mathbb{G}_n\phi_{h_{1,n}} \geq \sup_{h \in \mathcal{H}}\, \mathbb{G}_n\phi_{h} - \epsilon_n$ and $
 P_0(\phi_{h_{1,n}}-\phi_{h_{2,n}})^2 \leq \inf_{h \in \mathcal{H}_n} P_0(\phi_{h_{1,n}}-\phi_h)^2 + \epsilon_ n$.
Now, with probability tending to one,
\begin{align*}
\inf_{h \in \mathcal{H}} P_0(\phi_{h_{1,n}}-\phi_{h})^2 \leq \sup_{h_1 \in \mathcal{H}}\inf_{h_2 \in \mathcal{H}_n} P_0(\phi_{h_{1}}-\phi_{h_{2}})^2 \leq \epsilon_n\ ,
\end{align*}
and so, $P_0(\phi_{h_{1,n}}-\phi_{h_{2,n}})^2 = o_P(1)$. We can write $\sup_{h \in \mathcal{H}} \, \mathbb{G}_n\phi_{h} - \sup_{h \in \mathcal{H}_n} \, \mathbb{G}_n\phi_{h} =A_n+B_n+C_n$ with $A_n:=\sup_{h\in\mathcal{H}}\mathbb{G}_n(\phi_h-\phi_{h_{1,n}})$, $B_n:=\inf_{h\in\mathcal{H}_n}\mathbb{G}_n(\phi_{h_{2,n}}-\phi_h)$ and $C_n:=\mathbb{G}_n(\phi_{h_{1,n}}-\phi_{h_{2,n}})$.
By construction, we have that $0\leq A_n\leq \epsilon_n$, and so, $A_n=o_P(1)$. Additionally, we have that $B_n\leq 0$ because $h_{2,n} \in \mathcal{H}_n$.
Finally, since $P_0(\phi_{h_{1,n}}-\phi_{h_{2,n}})^2=o_P(1)$ and $\phi_{h_{1,n}}-\phi_{h_{2,n}}$ belongs to the Donsker class $\{\phi_{h_1}-\phi_{h_2}:h_1,h_2\in\bar{\mathcal{H}}\}$, we have that $C_n=o_P(1)$  by application of Lemma 19.24 of \cite{van2000asymptotic}.
Thus, we have established that $
\sup_{h \in \mathcal{H}} \, \mathbb{G}_n \phi_{h} - \sup_{h \in \mathcal{H}_n} \, \mathbb{G}_n \phi_{h}$ is bounded above by an $o_P(1)$ term.
A similar argument can be used to conclude the same of $
\sup_{h \in \mathcal{H}_n} \, \mathbb{G}_n \phi_{h} - \sup_{h \in \mathcal{H}} \, \mathbb{G}_n \phi_{h}$, thus establishing \eqref{Lemma1eq1}.

We now prove part (b). We note that
\begin{align*}
\left|\int_{\mathcal{H}_n}  [(P_n-P_0)\phi_h]^2 d\bar{Q}(h) - \int_{\mathcal{H}}  [(P_n-P_0)\phi_h]^2 d\bar{Q}(h) \right|\ &\leq\ \int_{(\mathcal{H}_n \cup \mathcal{H}) \setminus \left(\mathcal{H}_n \cap \mathcal{H}\right)}  [(P_n-P_0)\phi_h]^2 d\bar{Q}(h)\\
&\leq\ \bar{Q}\left(\left\{\mathcal{H}\cup\mathcal{H}_n\right\}\setminus\left\{\mathcal{H}\cap\mathcal{H}_n\right\}\right)\left[\sup_{h \in \bar{\mathcal{H}}}|(P_n-P_0)\phi_h|\right]^2.
\end{align*}
Since $\bar{\mathcal{H}}$ is a Donsker class and \eqref{QbarDist} holds by assumption, this implies that
\begin{align*}
\int_{\mathcal{H}_n}  [(P_n-P_0)\phi_h]^2 d\bar{Q}(h) = \int_{\mathcal{H}}  [(P_n-P_0)\phi_h]^2 d\bar{Q}(h) + o_P(n^{-1})\ ,
\end{align*}
and the result follows by an application of Slutsky's theorem and the continuous mapping theorem.

\end{document}